\documentclass[lettersize,journal]{IEEEtran}
\usepackage{amsmath,amsfonts}
\usepackage{algorithmic}
\usepackage{array}
\usepackage{textcomp}
\usepackage{stfloats}
\usepackage{url}
\usepackage{verbatim}
\usepackage{graphicx}
\usepackage{cite}
\usepackage[version=4]{mhchem}
\usepackage{chemformula}
\usepackage{multirow}
\usepackage{amsmath}
\usepackage[ruled,linesnumbered]{algorithm2e}
\usepackage{subfigure}  
\usepackage{bm}
\usepackage{booktabs}

\newcommand{\Rmnum}[1]{\uppercase\expandafter{\romannumeral #1}}

\def\BibTeX{{\rm B\kern-.05em{\sc i\kern-.025em b}\kern-.08em
    T\kern-.1667em\lower.7ex\hbox{E}\kern-.125emX}}
\usepackage{balance}
\begin{document}
\title{On-Site Precise Screening of SARS-CoV-2 Systems Using a Channel-Wise Attention-Based PLS-1D-CNN Model with Limited Infrared Signatures}
\author{Wenwen Zhang, Zhouzhuo Tang, Yingmei Feng,   Xia Yu, Qi Jie Wang, Zhiping Lin
\thanks{This work was partially supported by the National Key R\&D Program of China (2022YFE0102300), the National Medical Research Council (NMRC) under the Singapore-China joint grant on infectious diseases (MOH-000927), A*STAR grants (A2090b0144, M22K2c0080, and R23I0IR041), National Research Foundation grants (NRF-CRP29-2022-0003), and the National Natural Science Foundation of China under Grant 62203307. (\textit{Corresponding authors: \mbox{Qi Jie Wang}; Zhiping Lin.})  \par	
	Wenwen Zhang,  Qi Jie Wang and Zhiping Lin are with School of Electrical and Electronic Engineering, Nanyang Technological University, Singapore 639798, Singapore (e-mail: wenwen.zhang@ntu.edu.sg; qjwang@ntu.edu.sg; ezplin@ntu.edu.sg)\par
	Zhouzhuo Tang, Xia Yu are with School of Instrumentation and Optoelectronic Engineering, Beihang University, Beijing 100191, China (tangzhouzhuo@buaa.edu.cn; xiayu@buaa.edu.cn)\par
    Yingmei Feng is with Department of Science and Development, Beijing Youan Hospital, Capital Medical University, Beijing 100069, China (yingmeif13@ccmu.edu.cn)\par
    Wenwen Zhang and Zhouzhuo Tang contributed equally to this work.\par		  
}}

\markboth{IEEE TRANSACTIONS ON XXX}%
{How to Use the IEEEtran \LaTeX \ Templates}

\maketitle

\begin{abstract}
During the early stages of respiratory virus outbreaks, such as severe acute respiratory syndrome coronavirus 2 (SARS-CoV-2), the efficient utilize of limited nasopharyngeal swabs for rapid and accurate screening is crucial for public health. In this study, we present a methodology that integrates attenuated total reflection-Fourier transform infrared spectroscopy (ATR-FTIR) with the adaptive iteratively reweighted penalized least squares (airPLS) preprocessing algorithm and a channel-wise attention-based partial least squares one-dimensional convolutional neural network (PLS-1D-CNN) model, enabling accurate screening of infected individuals within 10 minutes. Two cohorts of nasopharyngeal swab samples, comprising 126 and 112 samples from suspected SARS-CoV-2 Omicron variant cases, were collected at Beijing You'an Hospital for verification. Given that ATR-FTIR spectra are highly sensitive to variations in experimental conditions, which can affect their quality, we propose a biomolecular importance (BMI) evaluation method to assess signal quality across different conditions, validated by comparing BMI with PLS-GBM and PLS-RF results. For the ATR-FTIR signals in cohort 2, which exhibited a higher BMI, airPLS was utilized for signal preprocessing, followed by the application of the channel-wise attention-based PLS-1D-CNN model for screening. The experimental results demonstrate that our model outperforms recently reported methods in the field of respiratory virus spectrum detection, achieving a recognition screening accuracy of 96.48\%, a sensitivity of 96.24\%, a specificity of 97.14\%, an F1-score of 96.12\%, and an AUC of 0.99. It meets the World Health Organization (WHO) recommended criteria for an acceptable product: sensitivity of 95.00\% or greater and specificity of 97.00\% or greater for testing prior SARS-CoV-2 infection in moderate to high volume scenarios.      
\end{abstract}


\section{Introduction}
\IEEEPARstart{R}espiratory viruses, such as severe acute respiratory syndrome coronavirus 2 (SARS-CoV-2), are frequently highly contagious, spread rapidly, and possess diverse transmission routes \cite{b1, b2}. The transmission routes of the SARS-CoV-2 virus primarily include respiratory, aerosol, and contact transmission \cite{b5, r3, r4, r5}.  Since the outbreak of (Coronavirus Disease 2019) COVID-19 in December 2019, a variety of techniques have been employed to screen individuals infected with coronavirus. These techniques include nucleic acid testing using methods such as reverse transcription-quantitative polymerase chain reaction (RT-qPCR), constant temperature amplification, sequencing, and clustered regularly interspaced short palindromic repeats (CRISPR) for nucleic acid detection \cite{ b7, r6}, as well as antigen testing \cite{b8}, antibody testing \cite{b9, r7}, lung computed tomography (CT) examination \cite{b10}, and routine blood examination \cite{b13}.\par
The RT-qPCR method, renowned for its exceptional specificity and sensitivity, stands as the gold standard for detecting SARS-CoV-2 \cite{b14, b15, b16, r9}. However, this approach is limited by the requirement for expensive qPCR instruments and skilled operators. Additionally, the intricate detection process contributes to a time-consuming procedure \cite{b17}, with current test result retrieval taking at least six hours, or even extending until the following day. Testing is confined to designated laboratories, with on-site real-time testing currently unavailable. 
Antigen self-testing serves as a valuable supplementary method for SARS-CoV-2 screening, effectively reducing the burden on medical resources \cite{b18, b19}. The production cost per unit of antigen self-test reagents is approximately \$0.415, and they are consumables that cannot be reused. This cost becomes significant when used for large-scale populations. Furthermore, antigen tests have the limitation of relatively low specificity and sensitivity.  Lung CT is considered the gold standard imaging modality for diagnosing COVID-19 and is a primary method used for screening patients with COVID-19 \cite{b20, b22}. While imaging diagnosis shows higher sensitivity than nucleic acid testing, it also relies on evaluation by a professional doctor and typically takes 6 to 8 hours to receive the report. The limited availability of CT equipment and a shortage of medical diagnostic professionals have made the widespread implementation of COVID-19 screening unfeasible \cite{b23}. To improve detection speed, many researchers now employ viral molecular spectra combined with deep learning models to identify potential SARS-CoV-2 infections. For example, Huang \textit{et al.} \cite{s34} employed surface-enhanced Raman spectroscopy on throat swabs with supervised deep learning to complete virus screening within 20 minutes.\par
To further explore the potential of attenuated total reflection-Fourier transform infrared spectroscopy (ATR-FTIR) spectroscopy for extensive and rapid screening of SARS-CoV-2 patients with a limited number of nasopharyngeal swabs available during the initial spread of the novel coronavirus variant, we designed an integrated system for on-site SARS-CoV-2 screening in near-real-time with reusability. This system directly captures ATR-FTIR spectral signals from pharyngeal swab samples, applies advanced signal preprocessing techniques such as adaptive iteratively reweighted penalized least squares (airPLS), and then inputs the processed data into the channel-wise attention-based partial least squares-one dimensional-convolutional neural network (PLS-1D-CNN) model to determine SARS-CoV-2 infection status. The main contributions of our work include:\par 
\begin{enumerate}[ 
	\setlength{\parindent}{2em}   ]
	\item The proposed biomolecular importance (BMI) evaluation method quantitatively assesses the significance of virus-related biomolecules in differentiating the quality of spectral signals collected under different experimental conditions. This approach elucidates the underlying biological correlations, thereby facilitating the selection of higher-quality spectra and standardizing experimental procedures to ensure consistent, high-quality spectral signal collection.\par	
	\item  The proposed PLS-1D-CNN model combines PLS's linear feature extraction with CNN's nonlinear learning, optimizing feature selection through channel-wise attention. PLS reduces spectral data dimensionality and extracts informative features, while the CNN captures nonlinear features. The channel attention mechanism dynamically adjusts channel importance, prioritizing relevant features for improved classification performance. Additionally, the model employs the airPLS preprocessing method to remove background noise and baseline drift in ATR-FTIR spectra, improving the signal-to-noise ratio and highlighting key absorption peaks in virus samples.\par
	\item Compared with RT-qPCR and antigen self-testing methods, the screening system using the channel-wise attention-based PLS-1D-CNN model offers repeatable, non-destructive screening with low inspection costs and high accuracy, sensitivity, and specificity. This model swiftly and accurately identifies SARS-CoV-2 infections and exhibits good convergence, making it suitable for early-stage epidemics when limited labeled samples are available.        \par
\end{enumerate}\par
The structure of the remaining sections is outlined as follows: Section II reviews related works. Section III describes the ATR-FTIR spectra acquisition protocol for nasopharyngeal swabs from patients suspected of SARS-CoV-2 infection. Section IV discusses band assignments for primary absorption peaks and the airPLS preprocessing for spectral baseline correction. Section V presents the channel-wise attention-based PLS-1D-CNN model, BMI evaluation method, and experimental results. Finally, Section VI summarizes the key conclusions of this study.\par
\section{Related work}
\subsection{contact-based methodology}
Currently, researchers and medical device technology companies worldwide are collaborating to develop a cost-effective, reusable diagnostic system capable of rapidly and accurately screening large populations on site. For instance, Tabrizi \textit{et al.} developed a novel, low-cost handheld impedimetric biosensor using bio-ready electrodes functionalized with SARS-CoV-2 antibodies for precise detection of SARS-CoV-2 infections \cite{b24}. Wang \textit{et al.} fabricated a low-cost planar fully depleted silicon-on-insulator (FDSOI) p-channel Schottky barrier (SB) metal-oxide-semiconductor field-effect transistor (MOSFET) for detecting virus ORFlab ribonucleic acid (RNA) genes \cite{b25}.  Quijano-Rubio \textit{et al.} developed protein-based biosensors to detect coronaviruses directly in patient samples without the need for genetic amplification. These biosensors produce light upon interacting with viral protein components or targeted COVID-19 antibodies, facilitating the detection of the SARS-CoV-2 virus \cite{b26}. Perdomo \textit{et al.} designed a portable bio-nanosensing electrode device that measures the electrochemical impedance spectra of a disposable, bio-modified screen-printed carbon-based working electrode (SPCE), detecting changes in the concentration of SARS-CoV-2 antigen molecules within a minute fluid sample on its surface for SARS-CoV-2 detection \cite{b27}. However, this device necessitates a sub-microliter fluid sample for testing. The process of obtaining and preparing such samples, particularly in non-laboratory settings, may present practical challenges and potentially affect the reliability of the results.       \par 
\subsection{non-contact-based methodology}
Differing from traditional contact-based tests, non-contact auscultation has garnered considerable interest among researchers \cite{b28, b29}. For instance, to address the challenges posed by PCR testing, which requires point-to-point sampling of nasopharyngeal swab samples, Li \textit{et al.} developed a compact and multifunctional pathogenic infection diagnosis system. This system not only diagnoses SARS-CoV-2 infection but also assesses symptom severity through breath and exhalation analysis \cite{b30}. Zhu \textit{et al.} proposed a non-contact approach that integrates modulation spectrum and linear prediction speech features, utilizing speech signals recorded at home with a portable device for detecting SARS-CoV-2 infections \cite{b31}. Similarly, Aytekin \textit{et al.} introduced a novel method employing a deep learning model called hierarchical spectrogram transformers to analyze recorded respiratory sounds for diagnosing coronavirus infections \cite{b32}. However, the effectiveness of the hierarchical spectrogram transformer (HST) model heavily relies on the quality and quantity of respiratory sound data. Insufficient or low-quality data can severely impact the model's performance and generalizability in practical measurements.       \par 

The advancement of molecular spectroscopy has paved a new path in diagnosing SARS-CoV-2 by identifying unique spectral patterns in viral molecules \cite{b33, b35, b36, c1, b37}. For instance, Ye \textit{et al.} from the University of California developed a portable virus capture device that utilized machine learning algorithms to differentiate between various viruses, including SARS-CoV-2, based on collected Raman spectroscopy \cite{b38}. Leong \textit{et al.} designed a portable breath analyzer using surface-enhanced Raman scattering technology, capable of swiftly identifying SARS-CoV-2 infections within less than 5 minutes \cite{b39}.  Besides leveraging the Raman spectrum of viruses, many researchers have focused on exploring the infrared spectrum of viruses. For instance, Wang \textit{et al.} investigated the use of Fourier transform infrared spectroscopy (FTIR) spectra from serum samples for rapid screening of the novel coronavirus using machine learning \cite{b41}. However, this method involves blood collection, which may cause discomfort for individuals. \par 

\section{Acquisition of ATR-FTIR Spectra from Nasopharyngeal Swabs in Suspected SARS-CoV-2 Cases}
\begin{figure*}[htbp]\centering
	\includegraphics[width=17.0cm]{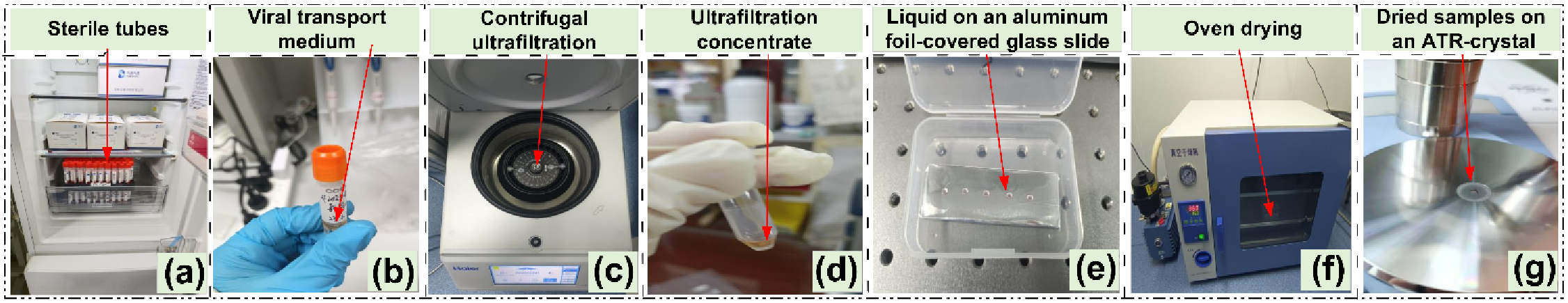}
	\caption{The primary experimental proceduces for collecting ATR-FTIR spectra of nansopharyngeal swabs.}
	\label{steps}
\end{figure*} 
This study received ethical approval from the Ethics Committee of Beijing Youan Hospital, Capital Medical University (approval number: \#2022-040). All research activities were conducted in accordance with the approved protocol and adhered to the principles of the China Food and Drug Administration/Good Clinical Practice (CFDA/GCP) and the Helsinki Declaration. In early 2023, two cohorts of nasopharyngeal swab samples were collected from patients potentially infected with the SARS-CoV-2 Omicron variant at Beijing Youan Hospital. Utilizing RT-qPCR as the gold standard, we categorized all swabs, assigning each sample a true positive or negative label. Cohort 1 consisted of 126 nasopharyngeal swabs, with 66 testing positive for COVID-19 and 60 testing negative. In cohort 2, 112 nasopharyngeal swabs were analyzed, with 53 samples confirmed positive and 59 confirmed negative. The primary experimental procedures for collecting ATR-FTIR spectra from nasopharyngeal swabs are illustrated in Fig. \ref{steps}. The detailed processes for obtaining ATR-FTIR spectra from pharyngeal swab samples in both cohorts are described below.\par
\begin{enumerate}[ 
	\setlength{\parindent}{2em}   ]
\item  Swab samples were stored in sterile tubes filled with viral transport medium (VTM) and placed in a freezer at -80 $^{\circ}\mathrm{C}$. These procedures are illustrated in Fig. \ref{steps} (a) and (b). Prior to collecting ATR-FTIR spectra from nasopharyngeal swabs, the swab samples were thawed and briefly vortexed.   \par
\item  For nasopharyngeal swabs in cohort 1, an aliquot of 4 $\mu$L of the solution was directly pipetted onto the ATR crystal and dried at room temperature for approximately 10 minutes. Alternatively, to expedite drying, approximately 10 $\mu$L of  each sample could be dispensed onto an aluminum foil-covered glass slide, as depicted in Fig. \ref{steps} (e), and then placed inside a drying box as shown in Fig. \ref{steps} (f) for about 2 to 3 minutes to ensure thorough drying.   \par
\item  For nasopharyngeal swabs in cohort 2, ultrafiltration centrifugation was employed to filter micromolecule compounds from VTM and concentrate biomolecules from the biological samples as shown in Fig. \ref{steps} (c). Initially, 400 $\mu$L of the samples were transferred into an ultrafiltration tube with a nominal molecular weight limit of 50 kDa and centrifuged at 14,000$\times$g for 5 minutes at 4 $^{\circ}\mathrm{C}$, resulting in a final volume of concentrated solution of approximately 30 $\mu$L. The concentrated solution, obtained after the separation of viruses and other biomacromolecules through ultrafiltration, is depicted in Fig. \ref{steps} (d). Subsequently, this concentrated solution underwent a drying process analogous to that described for cohort 1.  \par
\item  The ATR-FTIR spectral data were collected by placing aluminum foil with the dried sample on the crystal of a Bruker Alpha II FTIR spectrometer (Bruker Optics GmbH, Ettlingen, Germany), equipped with a platinum ATR module (Fig. \ref{steps} (g)). The spectral resolution was set to 2 cm\textsuperscript{-1}, capturing wave numbers from 4000 to 400 cm\textsuperscript{-1} over approximately 2 minutes, with 64 scans performed for both the background and samples.\par 

\item  After each measurement, the ATR crystal was cleaned with MilliQ water and ethanol. Each sample was measured three times, and the mean value was calculated as the spectral value, with any abnormal signals removed.  The entitle sample-to-result process can be completed within 10 minutes.
\end{enumerate}

\section{Main Absorption Peak Band Assignments and airPLS Baseline Correction}
\subsection{Infrared absorption bands and spectral interpretation} 
\begin{figure*}[htbp]
	\centering
	\subfigure[]{
		\begin{minipage}[t]{0.3\linewidth}
			\centering
			\includegraphics[ width=5.5cm,height=4.6cm]{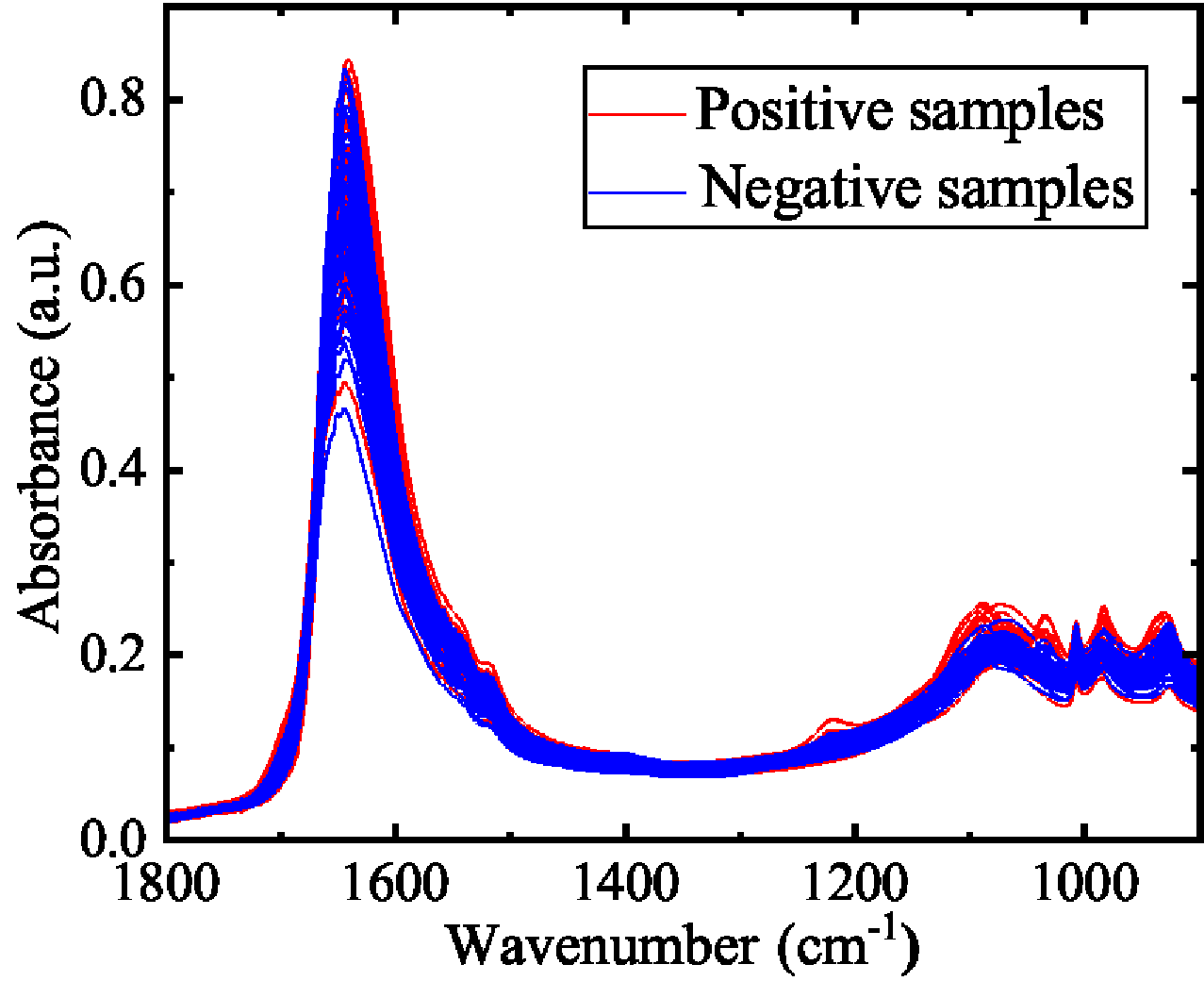}
		\end{minipage}%
	}\hspace{1pt}%
	\subfigure[]{ 
		\begin{minipage}[t]{0.60\linewidth}
			\centering
			\includegraphics[width=10.0cm,height=4.6cm]{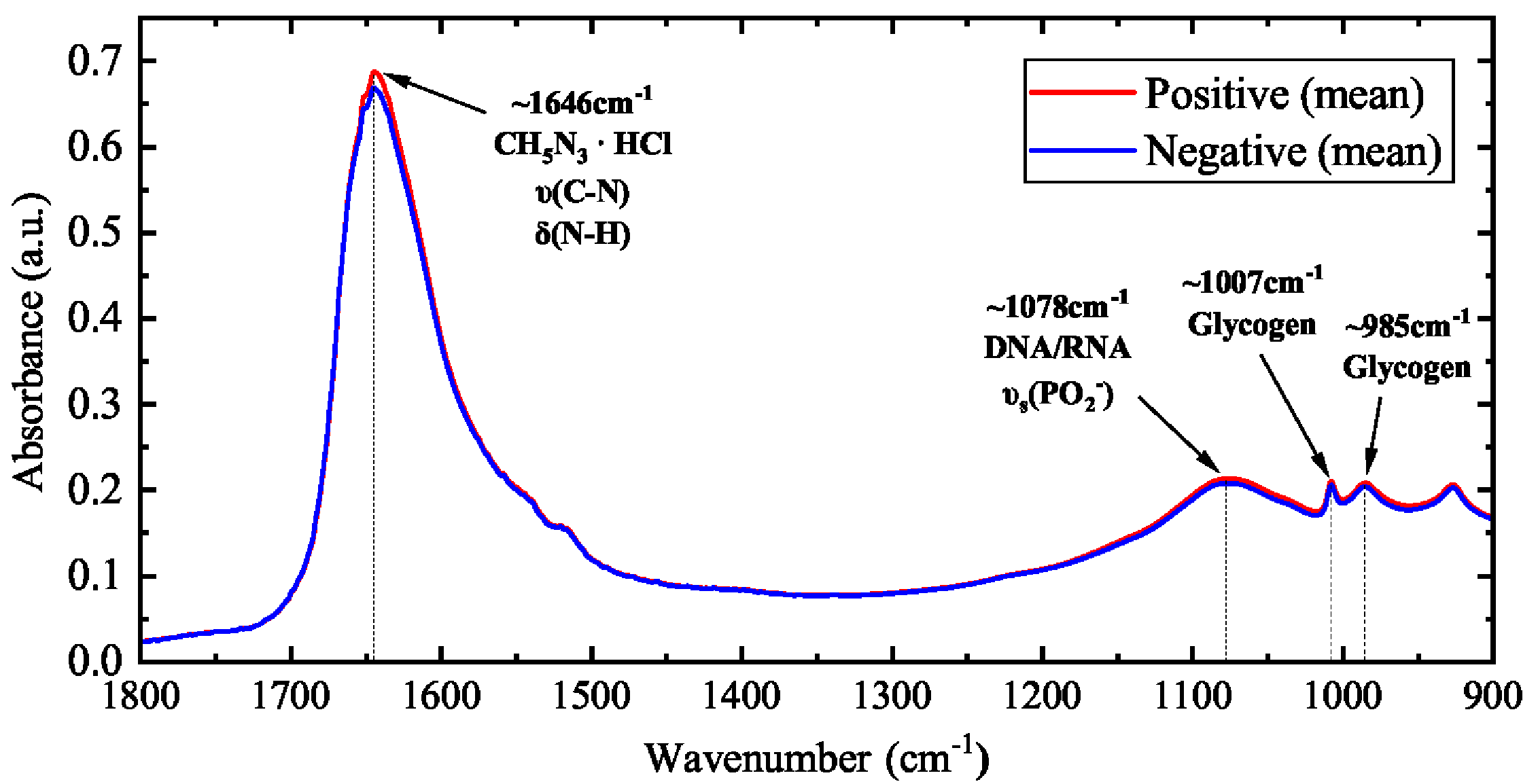}
		\end{minipage}%
	}%

	\subfigure[]{
		\begin{minipage}[t]{0.3\linewidth}
			\centering
			\includegraphics[ width=5.5cm,height=4.6cm]{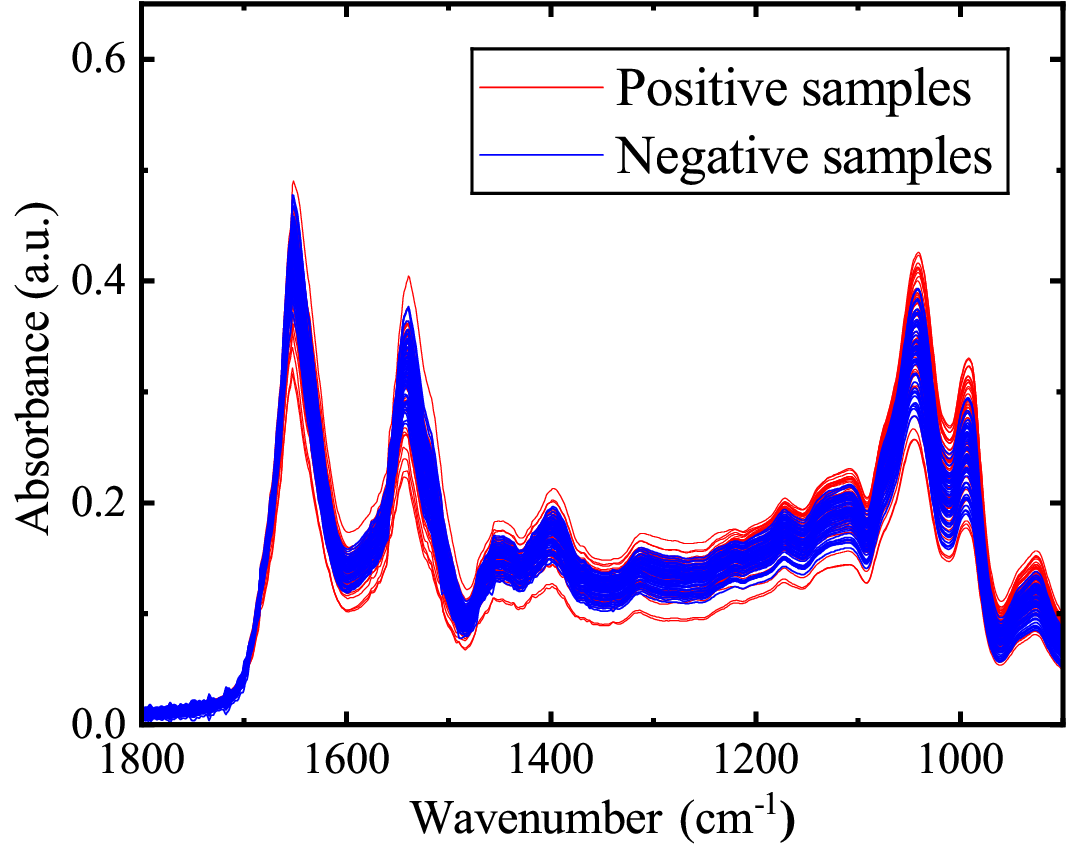}
		\end{minipage}%
	}\hspace{2pt}
	\subfigure[]{ 
		\begin{minipage}[t]{0.60\linewidth}
			\centering
			\includegraphics[width=10.0cm,height=4.6cm]{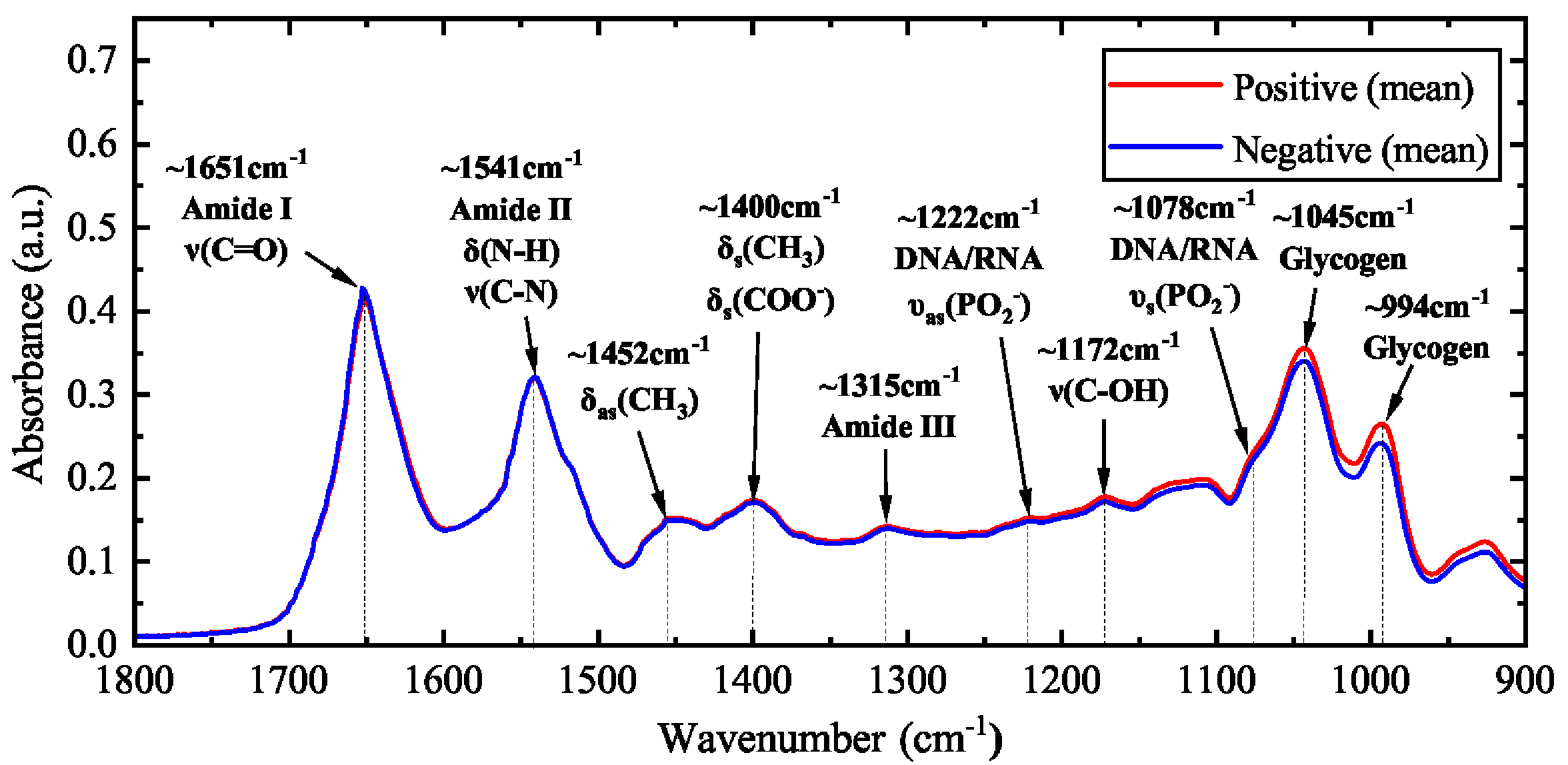}
		\end{minipage}%
	}%
	\caption{ The raw ATR-FTIR spectral signals of the pharyngeal swab samples in two cohorts. (a) cohort 1. (c) cohort 2. The mean ATR-FTIR spectral signals of positive and negative nasopharyngeal swabs two cohorts, along with annotations of the main absorption peaks. (b) cohort 1. (d) cohort 2.}
	\label{spectra}
\end{figure*}

The ATR-FTIR spectral signals derived from two cohorts of pharyngeal swab samples, spanning the truncated wavenumber range of 1800 to 900 cm\textsuperscript{-1} and recognized as the biofingerprint, are visualized in Fig. \ref{spectra} (a) and Fig. \ref{spectra} (c). The selected wavenumber range of 1800 to 900 cm\textsuperscript{-1} includes key biomarkers crucial for the primary classification of viral molecules, such as lipids (peak at 1750 cm\textsuperscript{-1} ({$\nu$(C=C)}), 1736 ($\nu$(C=O stretching))  ), amide \Rmnum{1} (peak at 1685  cm\textsuperscript{-1} (disorder structure-non-hydrogen  bonded), 1659 cm\textsuperscript{-1}), amide \Rmnum{2} (peak at 1549 cm\textsuperscript{-1}, 1517cm\textsuperscript{-1})  amide \Rmnum{3} (peak at 1307 cm\textsuperscript{-1}, 1260$\sim$1250  cm\textsuperscript{-1}), nucleic acids (peak at 1224 ($\nu$(asymmetric $\rm PO_{2}^{-}$ streching)), 1087 (peak at $\nu$(symmetric $\rm PO_{2}^{-}$ streching)) and carbohydrates (peak at 1150 (\ch{$\nu$ (C-O)}  stretching)) \cite{b42}. It is important to note that the infrared absorption range of specific molecular chemical bonds or functional groups can vary in different measurement conditions. Table \ref{bands} summarizes the primary biomolecules found in viruses, along with their respective absorption wavenumber bands and approximate absorbance peaks, as generalized as possible based on our experiments and the literature \cite{b42, b43, b44, b45, b46}. \par
\begin{table}[htbp]
	\centering
	\caption{The primary biomolecules of viruses and their principal infrared spectrum absorption bands}
	\scalebox{0.85}{
		\begin{tabular}{ccc}
			\hline \hline
			Biomolecules                    & Primary band (cm\textsuperscript{-1})                                           & Approximate peak (cm\textsuperscript{-1})                                                                                                              \\  \hline 
			Lipids 
			&                         
			\begin{tabular}[c]{@{}c@{}}1760$\sim$1720\\ 1470$\sim$1430\end{tabular}&    \begin{tabular}[c]{@{}c@{}} 1750 (\ch{$\nu$ (C = C)}), 1736 ($\nu$(C=O stretching))  \end{tabular}                                                    \\ 
			Amide \Rmnum{1}                         & 1700$\sim$1600                                                & \begin{tabular}[c]{@{}c@{}}1685 (disorder structure-non-hydrogen \\ bonded), 1659 \end{tabular}                                                                                       
			
			\\ 
			Amide \Rmnum{2}                        & 1600$\sim$1500                                                & 1549, 1517                                                                                                                    
			
			\\ 
			Amide \Rmnum{3}                       & 1350$\sim$1220                                                 & 1307, 1260$\sim$1250                                                                                                                         
			
			                                                             \\ 
			Nucleic acids     
			& \begin{tabular}[c]{@{}c@{}}1240$\sim$1220\\ 1120$\sim$1040\\ 1000$\sim$950\end{tabular} & \begin{tabular}[c]{@{}c@{}} 1224 ($\nu$ (asymmetric $\rm PO_{2}^{-}$ streching ))\\ 1087 ($\nu$ (symmetric $\rm PO_{2}^{-}$ streching))   \end{tabular} 
			
			\\ 
			\multicolumn{1}{l}{Carbohydrates}                   & \begin{tabular}[c]{@{}c@{}}1180$\sim$1100\\ 1050$\sim$970\end{tabular}&     \begin{tabular}[c]{@{}c@{}}1150 (\ch{$\nu$ (C-O)}  stretching) \\ 1050 (\ch{$\nu$ (C-O}  stretching coupled with \\ C-O bending of the C-OH)) \end{tabular}                                                                                  
			
			\\ \hline \hline
	\end{tabular}}
	\label{bands}
\end{table}

As depicted in Fig. \ref{spectra} (a) and Fig. \ref{spectra} (c), while the absorption spectrum curves for the positive and negative samples share a similar overall shape trend, they manifest differences in certain localized spectral bands. The mean spectra of positive and negative samples in cohort 1 and cohort 2, highlighting their main absorption peaks, are depicted in Fig. \ref{spectra} (b) and Fig. \ref{spectra} (d). The spectra of cohort 1 exhibit a notable maximum absorption peak at approximately 1646 cm\textsuperscript{-1}, attributable to the large quantity of guanidine salts, a common nucleic acid extraction reagent in VTM. The strong absorption peaks from VTM components mask the absorption peaks of biomolecules at ultra-low concentrations, particularly the amide bands of proteins. Consequently, only certain absorption peaks of nucleic acids and carbohydrates are visible in the 1100$\sim$900 cm\textsuperscript{-1} band.  \par
The spectra of concentrated samples after ultrafiltration in cohort 2 revealed the main infrared absorption peaks characteristic of typical biological samples. In accordance with Table \ref{bands}, Amide \Rmnum{1}, Amide \Rmnum{2}, Amide \Rmnum{3} and carbohydrates (\ch{$\nu$ (C-O}  stretching coupled with \ch{C-O} bending of the \ch{C-OH})) exhibit absorption peaks at  approximate wave numbers of  1651 cm\textsuperscript{-1}, 1541 cm\textsuperscript{-1}, 1315 cm\textsuperscript{-1}, 1045 cm\textsuperscript{-1} and 994 cm\textsuperscript{-1}, respectively, within the spectral curves.  Due to the inherently low levels of the free component in the swab samples and the presence of other substances in saliva, such as cell debris, bacteria, throat secretions, or traces of blood, lipids ($\nu${(C=C)}), DNA and RNA (\ch{$\nu$}(asymmetric \ch{$\rm PO_{2}^{-}$} stretching), \ch{$\nu$}(symmetric \ch{$\rm PO_{2}^{-}$} stretching)) did not exhibit maximum absorption peaks consistent with the experimental results recorded in Table \Rmnum{1}. Therefore, relying solely on the characteristic bands described in Table \Rmnum{1} as features to distinguish positive and negative nasopharyngeal swab samples is insufficient for achieving optimal recognition accuracy. 
Based on the analysis of the original spectral signal, it becomes evident that establishing a key biomolecule feature band for effective extraction is crucial to improving virus identification accuracy.\par 

\subsection{AirPLS preprocessing method for spectra baseline correction}
\begin{figure}[htbp]
	\centering
	\subfigure[]{ 
		\begin{minipage}[t]{0.5\linewidth}
			\centering
			\includegraphics[width=4.3cm]{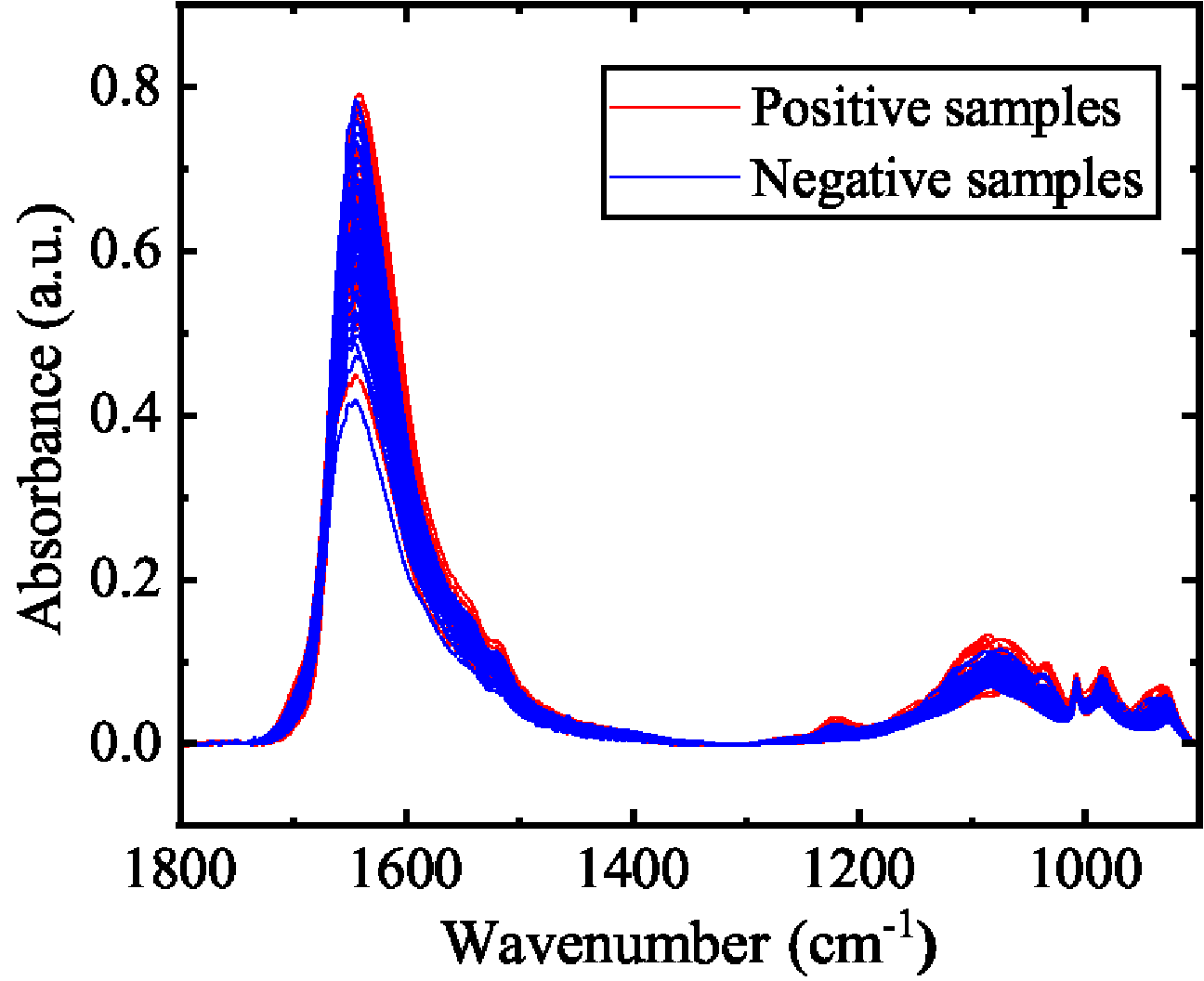}
		\end{minipage}%
	}%
	\subfigure[]{ 
		\begin{minipage}[t]{0.5\linewidth}
			\centering
			\includegraphics[width=4.3cm]{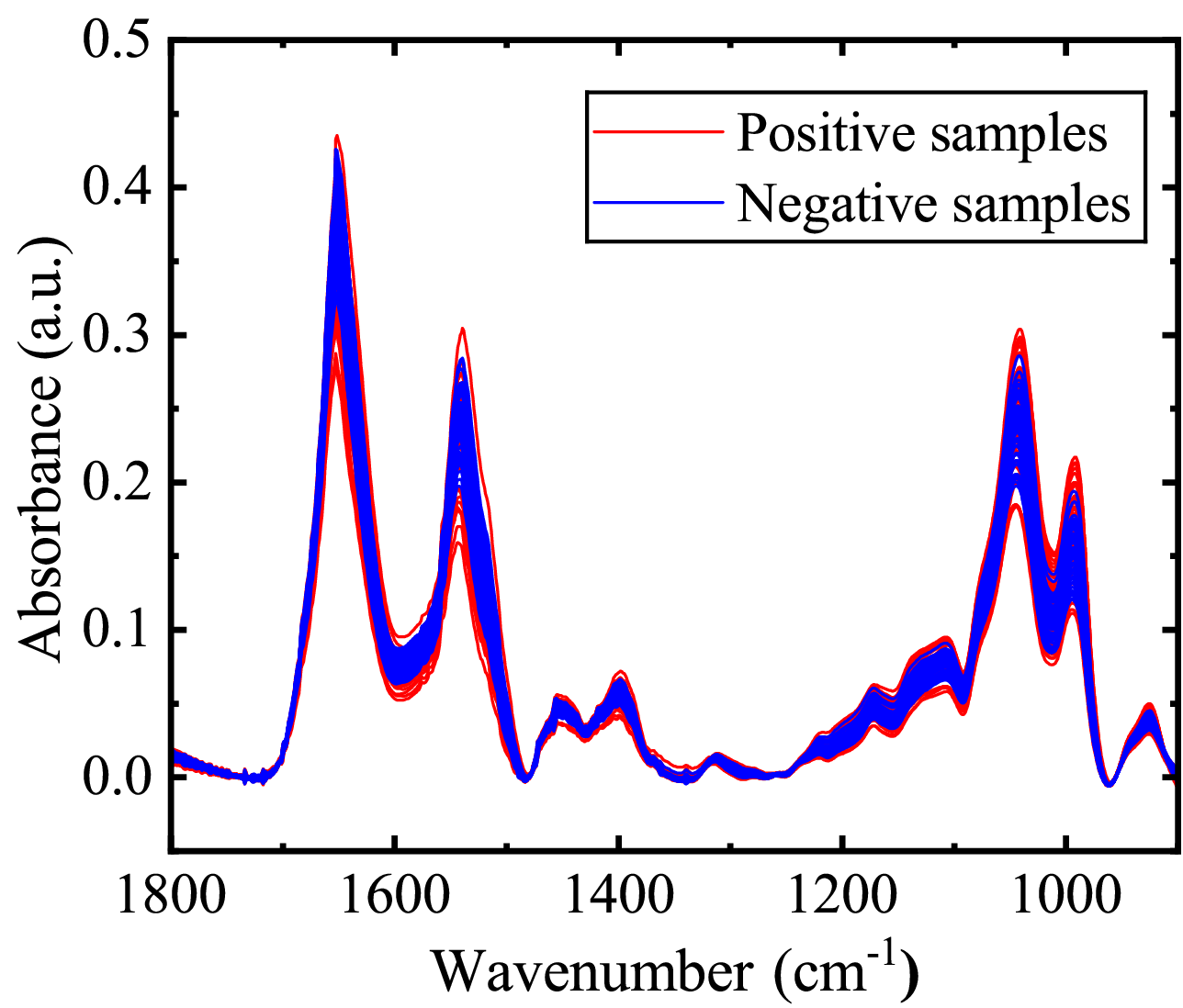}
		\end{minipage}%
	}%
	\caption{The spectral signal in two cohorts following baseline correction using the airPLS method. (a) cohort 1. (b) cohort 2.}
	\label{airpls}
\end{figure}

Prior to using raw spectra for analysis such as feature extraction and recognition, baseline correction is a critical step in preprocessing infrared absorption spectra, primarily employed to remove background signals arising from the absorption of substances not under analysis from the spectra. In our study, we utilize the airPLS method for performing baseline correction on ATR-FTIR spectra of both positive and negative samples. The raw spectra for cohort 1 and cohort 2, after applying the airPLS method for baseline correction, are illustrated in Fig. \ref{airpls} (a) and (b), respectively. After removing the background noise, the spectral signals show smoother profile, with the absorption intensity of spectral signals from both positive and negative samples being more concentrated. \par

\begin{figure}[htbp]\centering
	\includegraphics[width=8.8cm]{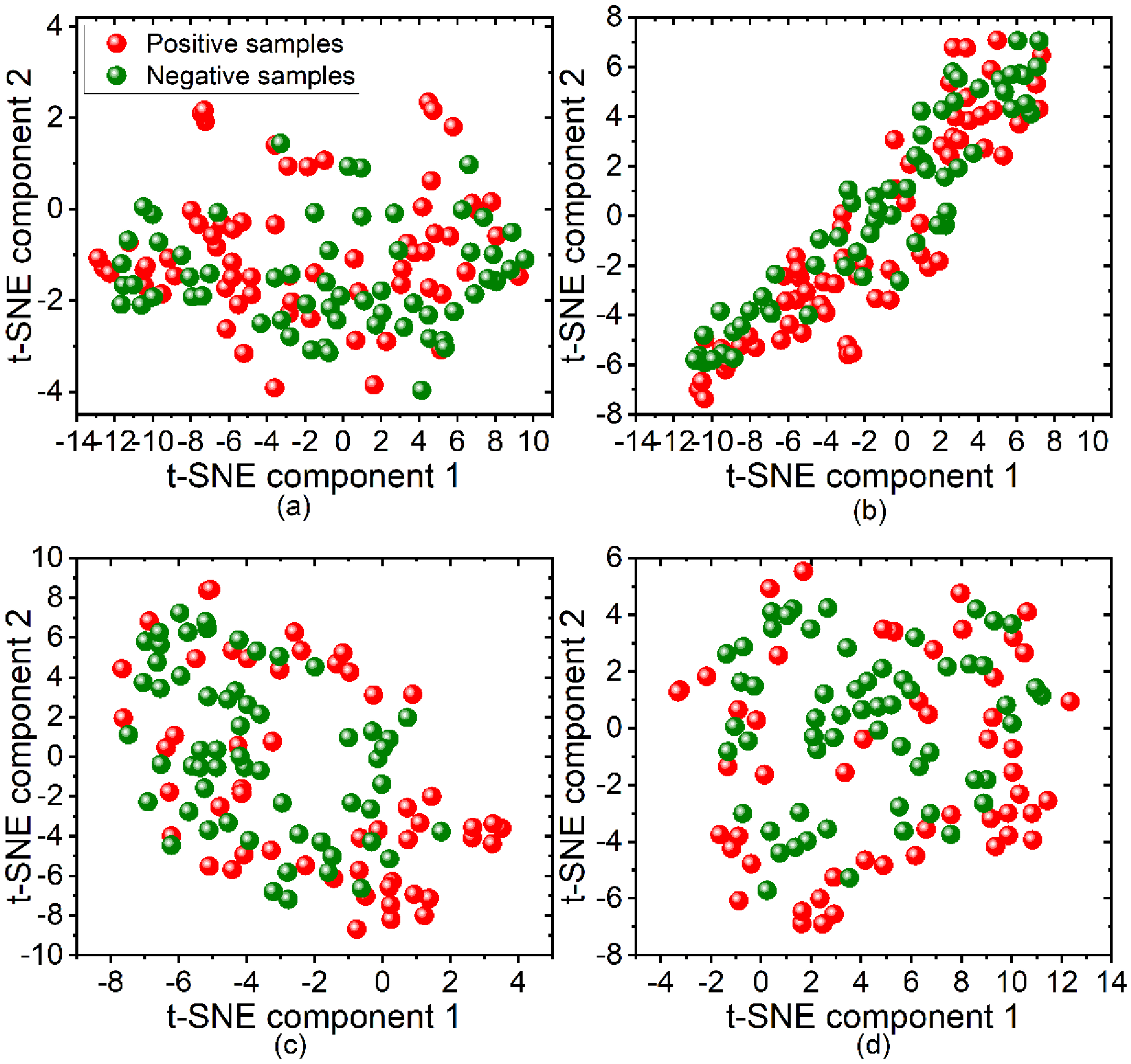}
	\caption{ (a) Dimensionality reduction of the original ATR-FTIR spectra using the t-SNE method for cohort 1. (b) Dimensionality reduction of airPLS baseline-corrected spectra  using the t-SNE method for cohort 1. (c) Dimensionality reduction of the original spectra using the t-SNE method for cohort 2. (d) Dimensionality reduction of airPLS baseline-corrected spectra using the t-SNE method for cohort 2. }
	\label{tsne_airPLS}
\end{figure}

To visually represent the separability of high-dimensional data samples, we use t-distributed stochastic neighbor embedding (t-SNE) to reduce the 874-dimensional spectral features to two dimensions. The distributions resulting from dimensionality reduction of the raw spectral signal and the airPLS baseline-corrected spectral signal for cohort 1 and cohort 2 are shown in Fig. \ref{tsne_airPLS} (a), (b), (c) and (d), respectively. Although the airPLS algorithm effectively corrects the spectral baseline and eliminates drift caused by background noise, sample surface characteristics, or other external factors, it does not significantly enhance the separability between positive and negative samples. This highlights the significance of extracting valuable feature information from infrared spectrum data of viruses to achieve high classification accuracy.  \par


\section{channel-wise attention-based PLS-1D-CNN model, BMI evaluation method, and experiments}
\subsection{Architecture of channel-wise attention-based PLS-1D-CNN model and Spectral Quality Evaluation of Two Cohorts Using the BMI Method}
We were inspired by the PLS algorithm's capability to analyze datasets with a sample size much smaller than the number of feature attributes, making it ideal for processing small-sample data. Additionally, the 1D-CNN model excels at automatically extracting detailed feature information from one-dimensional signals such as infrared spectra and audio signals. Consequently, combining both methods proves particularly effective for analyzing limited virus sample spectra, especially during the early stages. By leveraging the strengths of both approaches, we developed the channel-wise attention-based PLS-1D-CNN model for identifying infrared spectrum signals post-airPLS baseline correction. This model comprises the PLS submodel and the 1D-CNN with channel-wise attention submodel.\par
\begin{algorithm}[htbp]
	\footnotesize
	\caption{PLS algorithm for feature extraction of spectral signals from pharyngeal swabs}
	\KwIn{Training spectra $\bm{X}_{train}$; labels $\bm{Y}_{train}$; number of components $N$; test spectra $\bm{X}_{test}$; tolerance $\epsilon$}
	\KwOut{Scores for training spectra $\bm{T}_{train}$; scores for test spectra $\bm{T}_{test}$; weight matrix $\bm{W}_{L}$; regression coefficients $\bm{B}$}	
	Compute the mean of $\bm{X}_{train}$: $\bar{\bm{X}}_{train}$ \;
	Center $\bm{X}_{train}$ by subtracting $\bar{\bm{X}}_{train}$ from each row: $\hat{\bm{X}}_{train} \leftarrow \bm{X}_{train} - \bar{\bm{X}}_{train}$ \;
	Initialize $\bm{T}_{train}$, $\bm{P}$, $\bm{Q}$, $\bm{W}_{L}$ as zero matrices\;
	Set $\bm{X}_{residual} \leftarrow \hat{\bm{X}}_{train}$ and $\bm{Y}_{residual} \leftarrow \bm{Y}_{train}$\; 	
	
	\For {$i = 1$ $to$ $N$} {
		Initialize $\bm{u}$ as the first column of $\bm{Y}_{residual}$\;
		
		\Repeat{$\delta < \epsilon$} {
			$\bm{w} \leftarrow (\bm{X}_{residual}^T \bm{u}) / \Vert \bm{X}_{residual}^T \bm{u} \Vert$\;  
			$\bm{t} \leftarrow (\bm{X}_{residual} \bm{w}) / \Vert \bm{X}_{residual} \bm{w} \Vert$\;  
			$\bm{q} \leftarrow (\bm{Y}_{residual}^T \bm{t}) / \Vert \bm{Y}_{residual}^T \bm{t} \Vert$\;  
			$\bm{u}_{new} \leftarrow \bm{Y}_{residual} \bm{q}$\;
			$\delta \leftarrow \Vert \bm{u} - \bm{u}_{new} \Vert$\;
			$\bm{u} \leftarrow \bm{u}_{new}$\;
		}
		
		$\bm{T}_{train}[:, i] \leftarrow \bm{t}$\;
		$\bm{P}[:, i] \leftarrow (\bm{X}_{residual}^T \bm{t}) / (\bm{t}^T \bm{t})$\;
		$\bm{Q}[:, i] \leftarrow \bm{q}$\;
		$\bm{W}_{L}[:, i] \leftarrow \bm{w}$\;
		
		$\bm{X}_{residual} \leftarrow \bm{X}_{residual} - \bm{t} (\bm{P}[:, i])^T$\;
		$\bm{Y}_{residual} \leftarrow \bm{Y}_{residual} - \bm{t} \bm{q}^T$\;
	}     
	
	Center $\bm{X}_{test}$ using $\bar{\bm{X}}_{train}$\;
	$\hat{\bm{X}}_{test} \leftarrow \bm{X}_{test} - \bar{\bm{X}}_{train}$\;  
	Project test data onto PLS components: $\bm{T}_{test} \leftarrow \hat{\bm{X}}_{test} \bm{W}_{L}$\;
	
	Compute regression coefficients: $\bm{B} = \bm{W}_{L} (\bm{P}^T \bm{W}_{L})^{-1} \bm{Q}^T$\;
	
	\Return $\bm{T}_{train}$, $\bm{T}_{test}$, $\bm{W}_{L}$, $\bm{B}$\;
\end{algorithm}

\begin{table}[htbp]
	\centering
	\caption{Matrix Names, Symbols, and Dimensions}
	\scalebox{0.8}{ 
		\begin{tabular}{ccc}
			\toprule \toprule
			\textbf{Matrix Name} & \textbf{Symbol} & \textbf{Dimensions} \\
			\midrule
			Training spectra           & $\bm{X}_{train}$    & $\mathbb{R}^{P \times M} \rightarrow \mathbb{R}^{89 \times 874}$   \\
			Labels                     & $\bm{Y}_{train}$    & $\mathbb{R}^{P \times 1} \rightarrow \mathbb{R}^{89 \times 1}$     \\
			Test spectra               & $\bm{X}_{test}$     & $\mathbb{R}^{Q \times M} \rightarrow \mathbb{R}^{23 \times 874}$   \\
			Centered training spectra   & $\hat{\bm{X}}_{train}$ & $\mathbb{R}^{P \times M} \rightarrow \mathbb{R}^{89 \times 874}$   \\ 
			Centered test spectra       & $\hat{\bm{X}}_{test}$  & $\mathbb{R}^{Q \times M} \rightarrow \mathbb{R}^{23 \times 874}$  \\ 
			Training scores            & $\bm{T}_{train}$    & $\mathbb{R}^{P \times N} \rightarrow \mathbb{R}^{89 \times N}$     \\
			Test scores                & $\bm{T}_{test}$     & $\mathbb{R}^{Q \times N} \rightarrow \mathbb{R}^{23 \times N}$     \\
			Weight matrix              & $\bm{W}_{L}$        & $\mathbb{R}^{M \times N} \rightarrow \mathbb{R}^{874 \times N}$    \\
			Regression coefficients    & $\bm{B}$            & $\mathbb{R}^{M \times 1} \rightarrow \mathbb{R}^{874 \times 1}$      \\
			Mean training spectra  & $\bar{\bm{X}}_{train}$ & $\mathbb{R}^{1 \times M} \rightarrow \mathbb{R}^{1 \times 874}$   \\
			Residual spectra           & $\bm{X}_{residual}$  & $\mathbb{R}^{P \times M} \rightarrow \mathbb{R}^{89 \times 874}$   \\
			Residual labels            & $\bm{Y}_{residual}$  & $\mathbb{R}^{P \times 1} \rightarrow \mathbb{R}^{89 \times 1}$     \\
			Scores vector for the response variable                     & $\bm{u}$            & $\mathbb{R}^{P \times 1} \rightarrow \mathbb{R}^{89 \times 1}$     \\
			Weight vector              & $\bm{w}$            & $\mathbb{R}^{M \times 1} \rightarrow \mathbb{R}^{874 \times 1}$    \\
			Projection vector          & $\bm{t}$            & $\mathbb{R}^{P \times 1} \rightarrow \mathbb{R}^{89 \times 1}$     \\
			Regression vector          & $\bm{q}$            & $\mathbb{R}^{1 \times 1} \rightarrow \mathbb{R}^{1 \times 1}$      \\
			Updated scores vector for the response variable                 & $\bm{u}_{new}$      & $\mathbb{R}^{P \times 1} \rightarrow \mathbb{R}^{89 \times 1}$     \\
			Loading matrix           & $\bm{P}$            & $\mathbb{R}^{M \times N} \rightarrow \mathbb{R}^{874 \times N}$    \\
			Loading matrix for the response matrix             & $\bm{Q}$            & $\mathbb{R}^{1 \times N} \rightarrow \mathbb{R}^{1 \times N}$      \\
			Updated residual spectra   & $\bm{X}_{residual}$ & $\mathbb{R}^{P \times M} \rightarrow \mathbb{R}^{89 \times 874}$   \\
			Updated residual labels    & $\bm{Y}_{residual}$ & $\mathbb{R}^{P \times 1} \rightarrow \mathbb{R}^{89 \times 1}$     \\
			\bottomrule 	\bottomrule 
	\end{tabular}}
     \label{symbol}
\end{table}
\begin{figure*}[htbp]\centering
	\includegraphics[width=17.6cm]{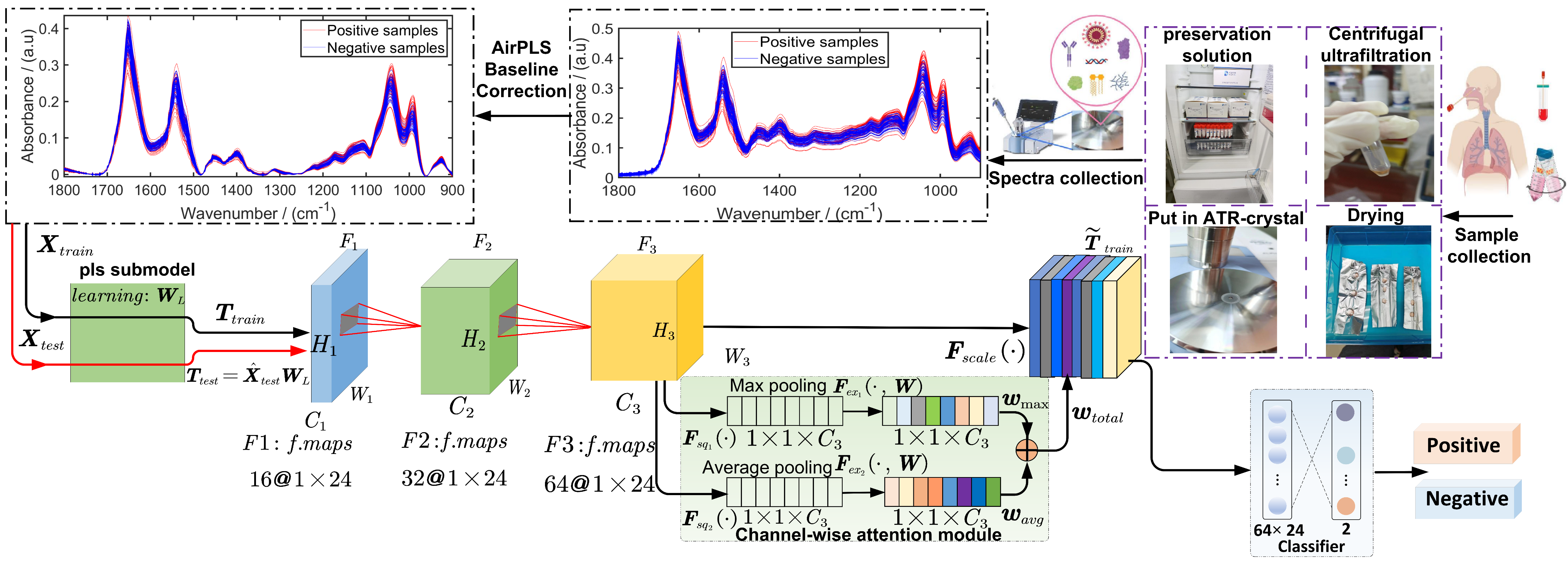}
	\caption{The channel-wise attention-based PLS-1D-CNN model for screening of infected individuals.}
	\label{model}
\end{figure*}
\begin{figure}[htbp]\centering
	\includegraphics[width=5.6cm]{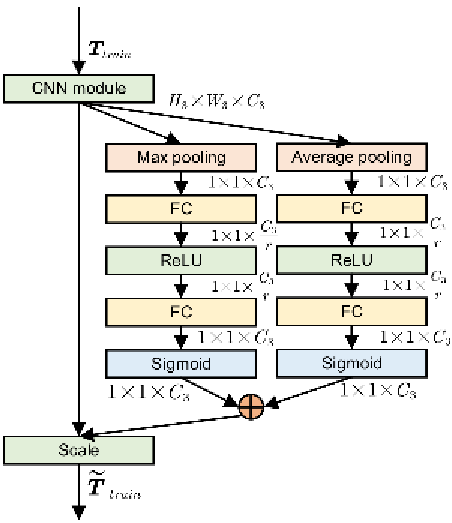}
	\caption{The detailed architecture of  channel-wise attention module.}
	\label{squeeze}
\end{figure} 
\subsubsection{PLS submodel} the PLS submodel serves as a supervised feature transformation and dimensionality reduction method. Its key advantage lies in its ability to transform features into a low-dimensional space, effectively addressing multicollinearity and high correlation in spectral data while preserving the most pertinent information for classification labels.  The 1D-CNN submodel with channel-wise attention is further employed for extracting higher-level features to enhance sample classification. The detailed architecture of model is depicted in Fig. \ref{model}. During the implementation of the channel-wise attention-based PLS-1D-CNN model,  we adopt a 5-fold cross-validation approach.  Initially, the spectral data sequence is shuffled and divided into five equal parts. Four of these parts (80\%) are employed to create the training dataset, denoted as $\bm{X}_{train} \in \mathbb{R}^{P \times M}$, while the corresponding labels are $\bm{Y}_{train} \in \mathbb{R}^{P \times 1}$, where $P$ and $M$ denote the number of training spectral samples and spectral features, respectively, with $P=89$ and $M=874$.  The remaining part (20\%) constitutes the test dataset, labeled as $\bm X_{test} \in \mathbb{R}^{Q \times M}$, with $Q=23$ and $M=874$. The pseudocode for implementing the PLS submodel, which transforms the features and reduces dimensionality from 874 to 24 principal components, is presented in Algorithm 1. The definitions and dimensions of the matrix symbols employed in the algorithm are provided in Table \ref{symbol}. For the variable $\epsilon$, we employ the default value of $10^{-10}$. The number of components, denoted as $N$, is set to 24.  The weight matrix, denoted as $\bm W_{L}$ in the PLS submodel, is acquired by learning from the labeled spectral signals within the training set. The spectral signal features in the training set are represented in the new feature space as $\bm {T}_{train}$, also known as the score matrix. The formula for calculating the representation of the test dataset in the new feature space, denoted as $\bm {T}_{test}$, is as follows:
\begin{equation}\label{f1}
\begin{aligned}
 \hat {\bm X}_{test} = \bm{X}_{test} - \bar {\bm X}_{train} 
\end{aligned}
\end{equation}

\begin{equation}\label{f2}
\begin{aligned}
\bm {T}_{test} =  \hat {\bm X}_{test} \bm{W}_{L}
\end{aligned}
\end{equation}
where $\bar {\bm X}_{train}$ denotes the mean matrix of the training set, and $\hat {\bm X}_{test}$ denotes the centered matrix of the test dataset. The projections $\bm {T}_{train}$ and $\bm {T}_{test}$, obtained by the PLS submodel in the new feature space with a feature length of 24, serve as the training and test datasets for the 1D-CNN with channel-wise attention submodel, respectively. \par

\subsubsection{Quantification of biomolecular importance map in PLS feature extraction submodel across two cohorts} 
To investigate the significance of specific spectral bands in the PLS submodel extracted spectral signal for effective classification, we introduced the BMI evaluation method to interpret, thereby facilitating the selection of higher-quality spectra and standardizes experimental procedures, ensuring the collection of consistent, high-quality spectral signals. \par
\begin{figure}[htbp]
	\centering
	\subfigure[]{ 
		\begin{minipage}[t]{0.5\linewidth}
			\centering
			\includegraphics[width=4.0cm]{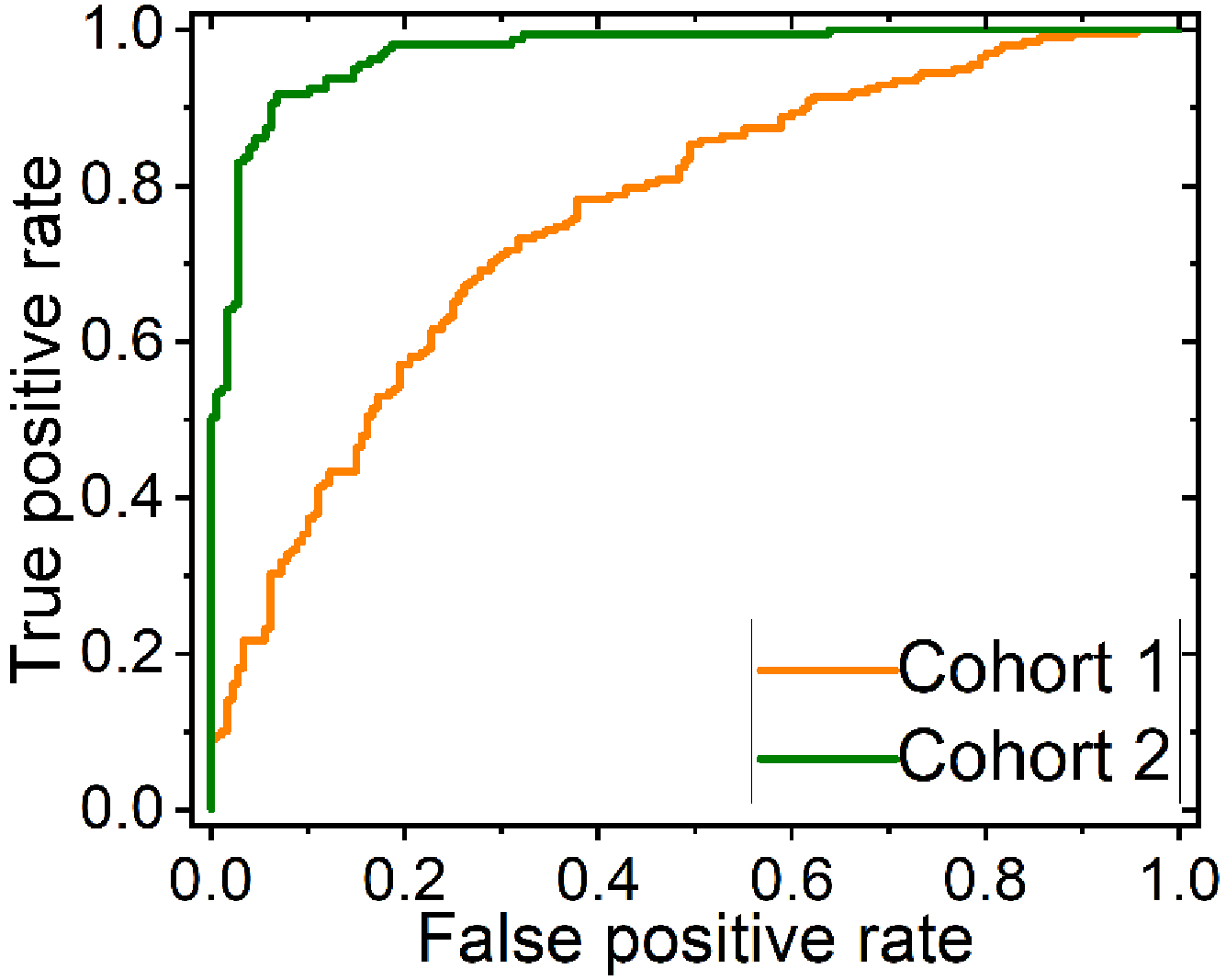}
		\end{minipage}%
	}%
	\subfigure[]{ 
		\begin{minipage}[t]{0.5\linewidth}
			\centering
			\includegraphics[width=4.0cm]{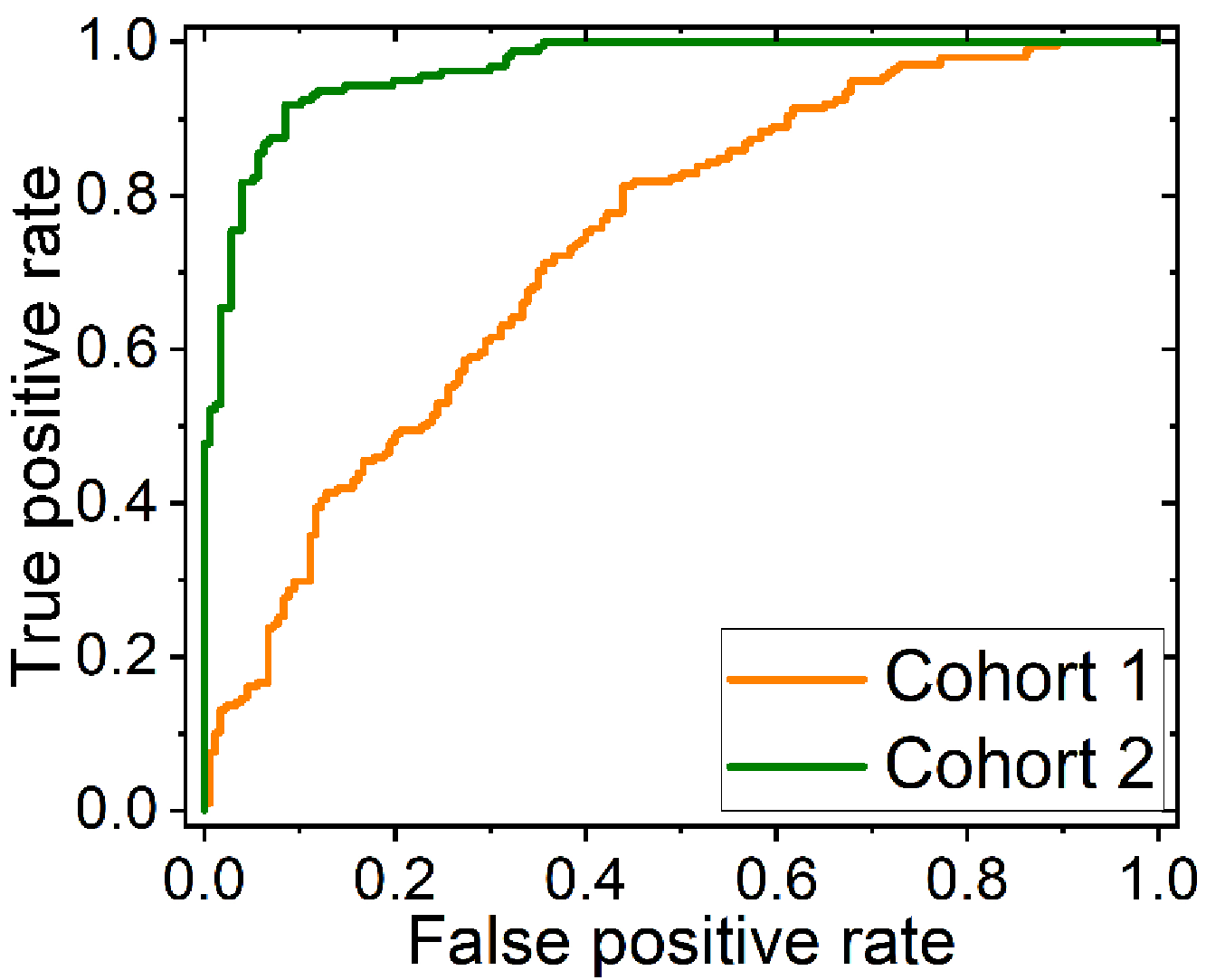}
		\end{minipage}%
	}%
	\caption{Comparison of ROC curves across two cohorts. (a) PLS-GBM. (b) PLS-RF.}
	\label{roc_cohort1_2}
\end{figure}

Intuitively, we can consider employing a unified feature extractor in conjunction with classification methods, such as partial least squares-random forest (PLS-RF)  and partial least squares-gradient boosting machine (PLS-GBM), to evaluate recognition accuracy, specificity, and sensitivity across both cohorts. By comparing these performance metrics, we were able to select higher-quality ATR-FTIR spectral signals from the two cohorts. The experimental results for accuracy, sensitivity, specificity, and F1-score obtained from the two cohorts are presented in Table \ref{performance}. Each metric was derived from the mean values obtained through five-fold cross-validation. PLS-RF achieved an accuracy of 91.37\% in distinguishing positive and negative nasopharyngeal swab spectra for cohort 2, while cohort 1 achieved only 67.78\%. Similarly, PLS-GBM achieved 91.05\% accuracy for cohort 2, compared to just 70.14\% for cohort 1. Moreover, the receiver operating characteristic (ROC) curves for each cohort are shown in Fig. \ref{roc_cohort1_2}. The area under the ROC curve (AUC) for PLS-GBM was 0.75 for cohort 1 and 0.97 for cohort 2, while for PLS-RF, the AUC values were 0.74 and 0.97, respectively. These results further highlight the superior diagnostic performance of the ATR-FTIR spectra in cohort 2. Although the consistency results of the above two methods demonstrate that the spectral quality of cohort 2 is higher than that of cohort 1, the factors driving this difference in accuracy remain unclear, particularly with respect to the biological basis of the feature extraction process employed for virus identification.\par

\begin{figure}[htbp]
	\centering
	\subfigure[]{ 
		\begin{minipage}[t]{0.5\linewidth}
			\centering
			\includegraphics[width=4.2cm,height=3cm]{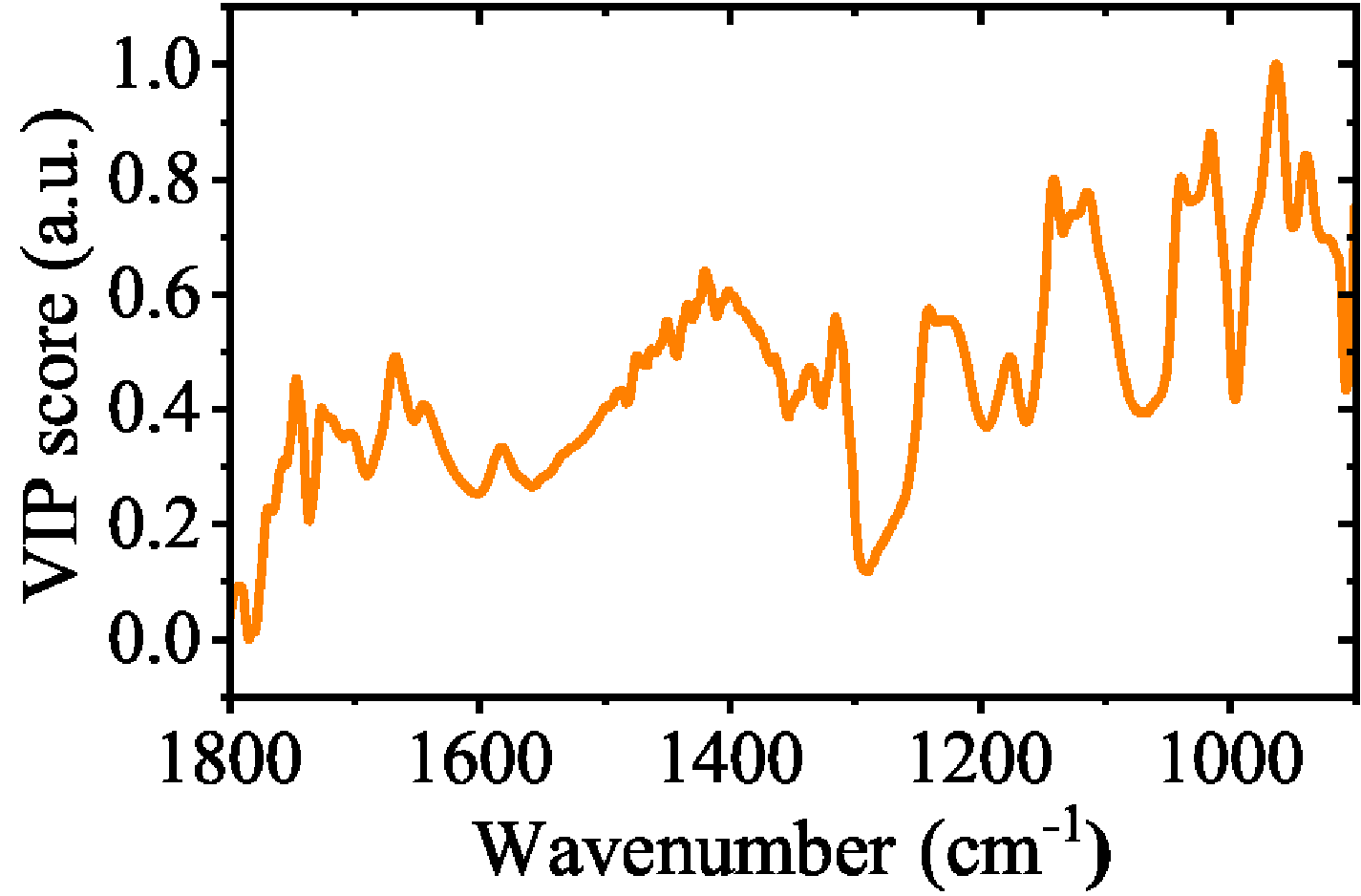}
		\end{minipage}%
	}%
	\subfigure[]{ 
		\begin{minipage}[t]{0.5\linewidth}
			\centering
			\includegraphics[width=4.2cm,height=3cm]{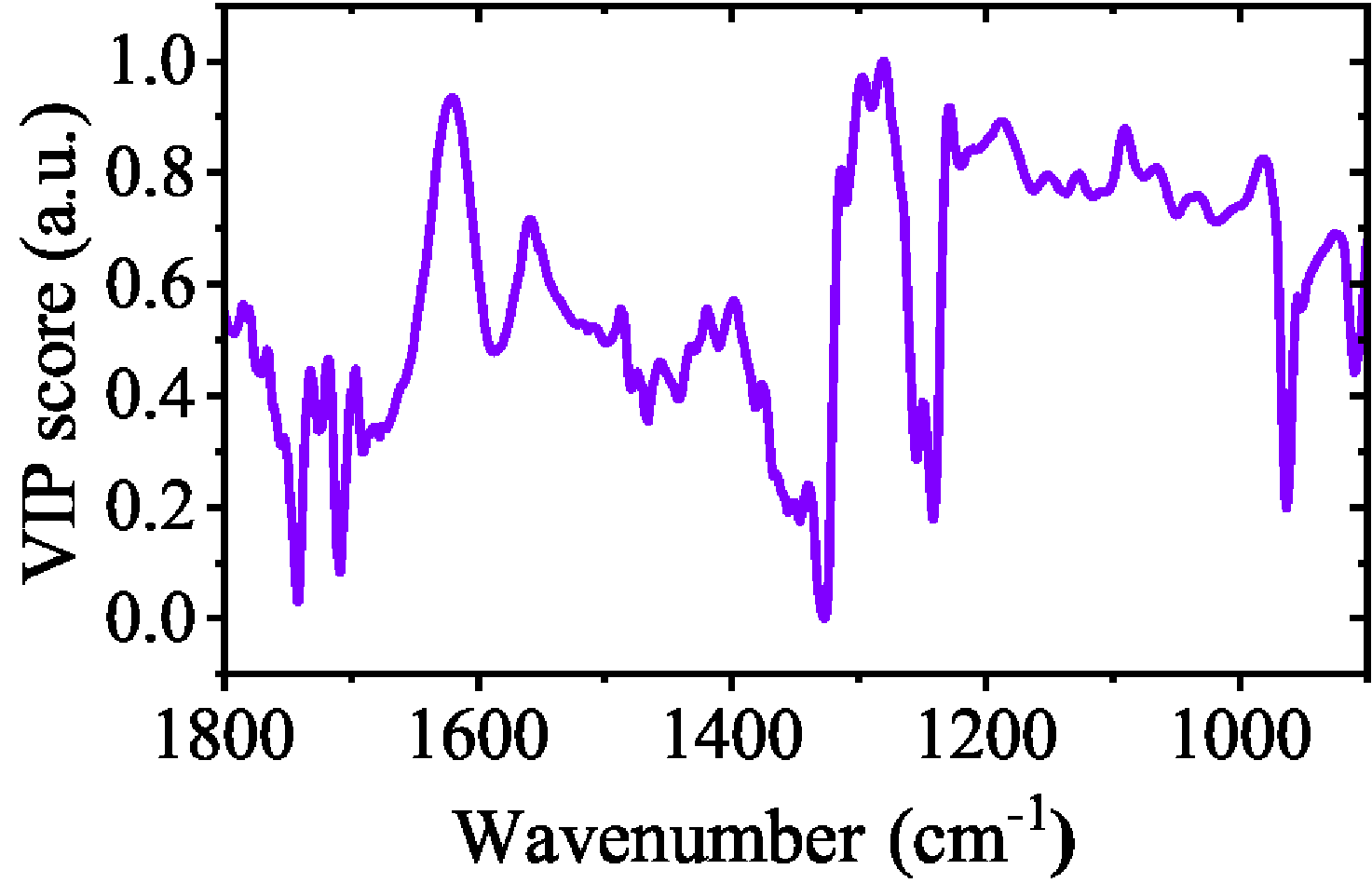}
		\end{minipage}%
	}%
	\caption{The VIP score curves for evaluating the significance of each wavenumber in the PLS-GBM and PLS-RF model across two cohorts. (a) cohort 1. (b) cohort 2.}
	\label{vipscore}
\end{figure}

\begin{table}[h]
	\centering
	\caption{Comparison of diagnostic performance between PLS-GBM and PLS-RF across two cohorts}
	\begin{tabular}{ccccc}
		\toprule \toprule
		& \multicolumn{2}{c}{\textbf{GBM}} & \multicolumn{2}{c}{\textbf{RF}} \\
		\cmidrule(lr){2-3} \cmidrule(lr){4-5}
		Dataset              & cohort 1        & cohort 2       & cohort 1       & cohort 2       \\
		\midrule
		\textbf{Accuracy (\%)}    & 70.14           & 91.05          & 67.78          & 91.37          \\
		\textbf{Sensitivity (\%)} & 73.63           & 87.6           & 74.42          & 91.39          \\
		\textbf{Specificity (\%)} & 69.17           & 93.68          & 64.61          & 92.08          \\
		\textbf{F1-score (\%)}    & 71.34           & 89.74          & 70.24          & 90.33          \\
		\bottomrule \bottomrule
	\end{tabular}
	\label{performance}
\end{table}
Immune responses to SARS-CoV-2 infection and the presence of the virus induce specific alterations in biomolecular content, resulting in subtle differences between the infrared spectra of positive and negative samples. However, other factors, such as variations in experimental conditions and the presence of additional substances, particularly components of viral transport media (VTM), can also introduce spectral differences and complicate classification. The reliability of the spectral dataset significantly impacts a classification model's ability to extract features associated with biomolecular absorption signatures. Consequently, we propose an evaluation metric, termed BMI, to quantitatively assess the significance of virus-related biomolecules in differentiating sample types, thereby elucidating the underlying biological correlations. The pseudocode for calculating the BMI index from the PLS submodel is provided in Algorithm 2. Variable Importance in Projection (VIP) scores were utilized to evaluate the significance of each wavenumber in the PLS-GBM and PLS-RF models. The formula for calculating VIP scores, based on the variance explained by each PLS component, is as follows:
\begin{equation}\label{f1}
\begin{aligned}
\bm{V}(j) = \sqrt{M\frac{\sum_{i=1}^N [\bm{SS}(i){(\bm{W}_{L}(i,j)/\|\bm{W}_{L}(i,:)\|)}^2]}{\sum_{i=1}^N\bm{SS}(i)}}
\end{aligned}	
\end{equation}
where $M$ and $N$ denote the number of spectral features and  components, respectively. The term ${(\bm{W}_{L}(i,j)/\|\bm{W}_{L}(i,:)\|)}^2$ reflects the importance of the $j$-th variable, where $\bm{W}_{L}$ denotes the weight matrix computed from Algorithm 1 (refer to Table \ref{symbol}). Additionally, $\bm{SS}(i)$ denotes the sum of squares explained by the $i$-th component, which can be computed as follows:\par
\begin{equation}\label{f1}
\begin{aligned}
 \bm{SS}(i)=\bm{Q}[:, i]^2\bm{T}_{train}[:, i]^T\bm{T}_{train}[:, i]  
\end{aligned}	
\end{equation}
where $\bm{Q}$ and $\bm{T}_{train}$ denote the loading matrix for the response variable and the training scores, respectively (refer to Table \ref{symbol}).
As shown in Fig. \ref{vipscore} (a) and (b), we normalized the VIP scores to a range of 0 to 1 to ensure a fair comparison between the two cohorts under consistent criteria.
Biomolecules, including lipids, proteins, carbohydrates, and nucleic acids, have potential infrared absorption bands, referred to as biomolecular ranges, which are presented in Table \Rmnum{1}. The BMI for a specific biomolecule is calculated as the root mean square (RMS) of the VIP scores corresponding to the wavenumbers within these biomolecular ranges. The calculation formula is as follows:
\begin{equation}\label{f1}
\begin{aligned}
\bm{\mathrm{BMI}} = \mathrm{RMS}({\bm{\mathrm{VIP}}}_{{\nu}\in{\bm{S}_{\mathrm{BR}}}})
=\sqrt{\frac{1}{K}\sum_{{\nu}\in{\bm{S}_{\mathrm{BR}}}} {{\bm{V}[\nu]}^2}}
\end{aligned}
\end{equation}
where $\nu$ denotes the variable of wavenumber, while $S_{\mathrm{BR}}$ denotes the set of wavenumbers within the biomolecular ranges, with $K$ indicating the cardinality of the set $S_{\mathrm{BR}}$. The BMI ranges from 0 to 1, with higher values signifying a greater contribution of the biomolecule to the spectral classification between SARS-CoV-2 infected and non-infected samples.\par
\begin{figure}[htbp]\centering
	\includegraphics[width=8.8cm]{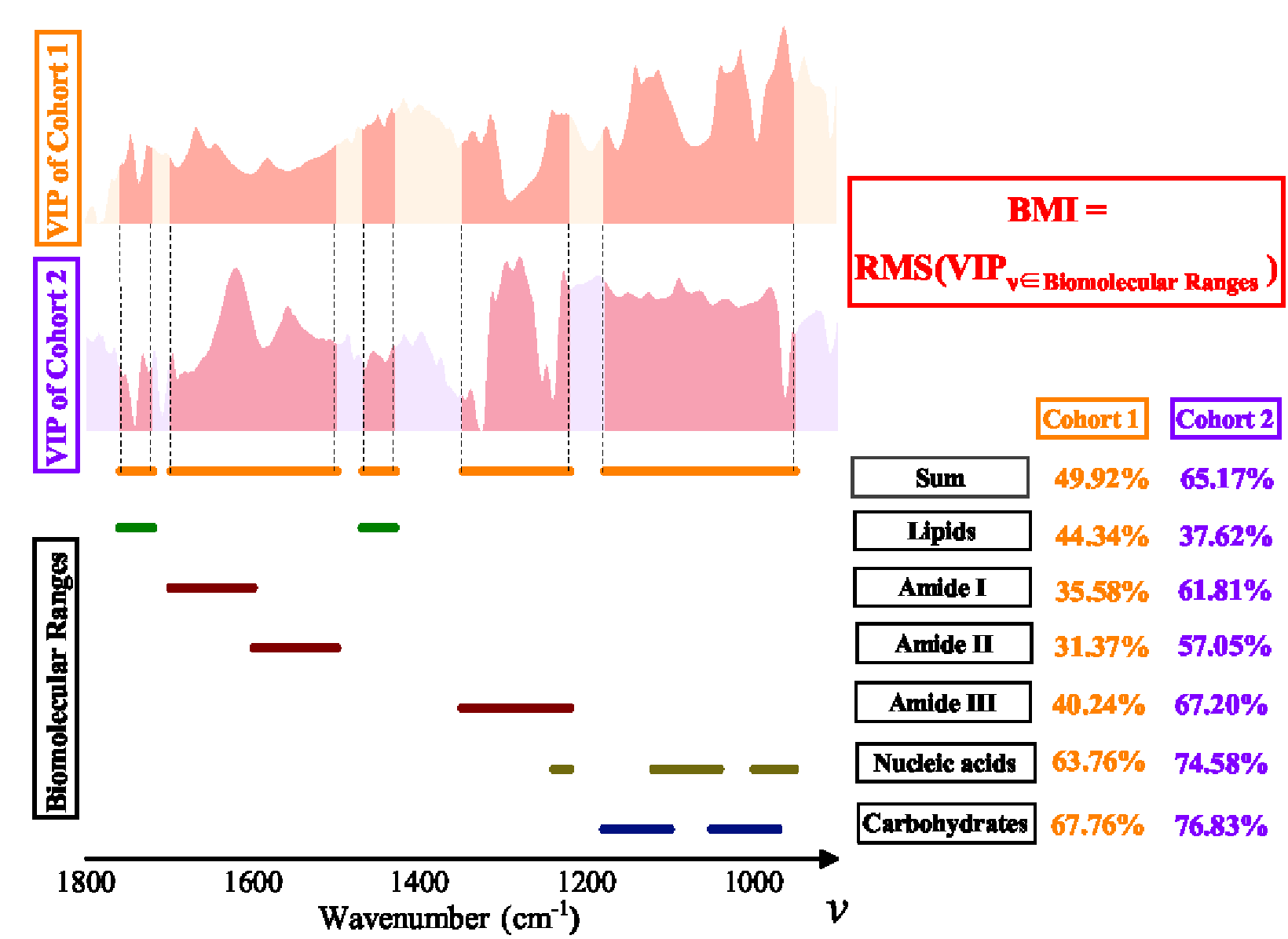}
	\caption{ Illustration of the quantification metric BMI of various biomolecules to evaluate the importance in sample distinguish across two cohorts.}
	\label{BMI}
\end{figure}

\begin{algorithm}[htbp]
	\footnotesize
	\caption{Algorithm for deriving BMI index from PLS}
	\KwIn{Weight matrix $\bm{W}_{L}$; training scores $\bm{T}_{train}$; loading matrix for the response matrix $\bm{Q}$; number of spectral features $M$; number of components $N$; set of wavenumbers within the biomolecular ranges $\bm{S}_{\mathrm{BR}}$; set cardinality $K$} 
	\KwOut{Variable importance of projection scores $\bm{V}$; biomolecular importance $\bm{\mathrm{BMI}}$}
	Compute the sum of squares explained by the $i$-th component:
	
	\For {$i = 1$ $to$ $N$} {
		$\bm{SS}(i) \leftarrow \bm{Q}[:, i]^2\bm{T}_{train}[:, i]^T\bm{T}_{train}[:, i]$\;
	}     
	
	Compute variable importance of projection scores:
	
	\For {$j = 1$ $to$ $M$} {
		$\bm{V}(j) \leftarrow \sqrt{M\frac{\sum_{i=1}^N [\bm{SS}(i){(\bm{W}_{L}(i,j)/\|\bm{W}_{L}(i,:)\|)}^2]}{\sum_{i=1}^N\bm{SS}(i)}}$\;
	}   
	
	Compute biomolecular importance:
	$\bm{\mathrm{BMI}} \leftarrow \sqrt{\frac{1}{K}\sum_{{\nu}\in{\bm{S}_{\mathrm{BR}}}} {{\bm{V}[\nu]}^2}}$
	
	\Return $\bm{V}$,  $\bm{\mathrm{BMI}}$;
\end{algorithm}
We calculated the Biomolecular Index (BMI) values for biomolecules across two cohorts, as illustrated in Fig. \ref{BMI}. The BMI values for most biomolecules in cohort 2 are significantly higher than those in cohort 1, with the exception of lipids. This finding indicates that PLS submodel extracts more features from the biomolecular absorption signatures in cohort 2. Notably, the three amide bands of proteins exhibit the most considerable difference between the two cohorts, likely due to the interference caused by high salt concentrations from viral transport medium (VTM) in the non-ultrafiltered cohort 1, which results in substantial overlap of absorption peaks with the amide bands. In contrast, the ultrafiltration procedure applied in cohort 2 facilitates the identification of proteins at higher abundances in the spectra, providing more insightful information. Proteins present in nasopharyngeal swabs, such as the SARS-CoV-2 spike glycoprotein (S protein), mucoprotein, albumin, and immunoglobulin G (IgG), demonstrate changes in both content and secondary structure in response to viral infection, thereby enhancing the distinction between the spectra of infected and non-infected samples \cite{b46, b47}. The contributions of nucleic acids and carbohydrates are likely attributable to SARS-CoV-2 RNA and highly glycosylated S proteins \cite{d2}. Compared to other biomolecules, lipids display lower BMI values in both cohorts (44.34\% in cohort 1 and 37.62\% in cohort 2). This aligns with previous studies that observed no significant absorbance changes in lipids \cite{b45}.\par
Based on the above analysis, we selected the spectra from cohort 2, which exhibited higher BMI values for biomolecules in the features extracted by the PLS submodel, as the focus for subsequent model analysis.  \par
\subsubsection{1D-CNN with channel-wise attention submodel}the widely used 2D-CNN is not suitable for direct application to one-dimensional signals, such as ATR-FTIR spectral signals. Therefore, we designed a 1D-CNN with channel-wise attention submodel, as depicted in Fig. \ref{model},  to identify positive and negative spectral signals.  It consists of three convolutional layers, one channel-wise attention module, and two fully connected layers. The experimental parameters for each convolutional layers are as follows: 1) input spectral signal size: 1$\times$24; number of input channels: 1. 2) First convolutional layer: kernel length: 3; stride: 1; number of ouput channels: 16; padding: same (to keep the input and output sizes consistent); and the size of the feature map output by this layer: 16@1$\times$24. 3) Second convolutional layer: kernel length: 3; stride: 1; number of ouput channels: 32; padding: same; and the size of the feature map output by this layer: 32@1$\times$24.  4)  Third convolutional layer: kernel length: 3; stride: 1; number of ouput channels: 64; padding: same; and the size of the feature map output by this layer: 64@1$\times$24. The output feature maps of 64 different channels, each with a length of 24, are fed into the channel-wise attention module.  The $\tilde {\bm{T}}_{train}$ feature maps, now weighted by channel, are then passed as input to the fully connected network in the subsequent classifier.  The detailed architecture of channel-wise attention module are depicted in Fig. \ref{squeeze}. The detailed implementation process is as follow: 1) squeeze operation: The spectral feature map output by the aforementioned CNN module, with dimensions of $ H_{3}\times W_{3}\times C\textsubscript{3}=1\times24\times64$, is pooled on each channel using both maximum pooling and average pooling, where $H_3$ and $W_3$ denote the dimensions of each feature layer, representing its length and width, while $C_3$ denotes the number of feature channels.   The resulting data vector for the max pooling operation, denoted as $\bm{v}_{\text{max}}$, and the average pooling operation, denoted as $\bm{v}_{\text{avg}}$, both have a size of 1$\times$1$\times$64, the expression for calculating the element in the $i$-th channel is as follows:\par

\begin{equation}\label{f1}
\begin{aligned}
v_{avg_i} = \bm{F}_{sq_1}(\bm{u}_i)=\frac{1}{W}\sum_{j=1}^{W}u_{i}(j)
\end{aligned}
\end{equation}

\begin{equation}\label{f1}
\begin{aligned}
v_{max_i} = \bm{F}_{sq_2} (\bm {u}_i) =\max\limits_{j=1, \cdots, W }{u_i(j)}
\end{aligned}
\end{equation}
where  $\bm {u}_i$ denotes the feature vector of the $i$-th channel output by the CNN module, $u_i(j)$ denotes its $j$-th element (where $j=1,2, \cdots, W$, and $W=W_3=24$). The purpose of the squeeze operation is to embed the global information of each channel in the feature map.  2) Excitation operation: to limit model complexity and improve generalization, it comprises two fully connected layers. The first fully connected layer achieves dimensionality reduction with a reduction ratio $r$, set to 16,  while the second fully connected layer increases dimensionality by the same ratio $r$. Then, the sigmoid function is applied to obtain the weight value for each channel. The expressions for calculating the weight vector $\bm{w}_{avg}$ from average pooling and  $\bm{w}_{max}$ from maximum pooling are as follows:\par  
\begin{equation}\label{f1}
\begin{aligned}
	\bm w_{avg}=\bm{F}_{ex_1}(\bm{v}, \bm{W}_e)=\sigma(\bm{W}_{e_2}\delta(\bm{W}_{e_1}\bm{v}_{avg}))
\end{aligned}
\end{equation}

\begin{equation}\label{f1}
\begin{aligned}
\bm w_{max}=\bm{F}_{ex_2}(\bm{v}, \bm{W}_e)=\sigma(\bm{W}_{e_2}\delta(\bm{W}_{e_1}\bm{v}_{max}))
\end{aligned}
\end{equation}
where $\delta$ denotes the ReLU activation function, and $\sigma$ denotes the sigmoid activation function. The matrices $\bm{W}_{e_1} \in \mathbb{R}^{\frac{C_3}{r} \times C_3}$ and $\bm{W}_{e_2} \in \mathbb{R}^{C_3 \times \frac{C_3}{r}}$ denote the weight matrices of the first and second fully connected layers, respectively. The vector $\bm{v}$, obtained through the aforementioned squeeze operation, corresponds to either the maximum pooling method ($\bm{v}_{max}$) or the average pooling method ($\bm{v}_{avg}$).  $\bm{W}_e$ denotes the weight coefficient matrix. Subsequently, the previously acquired weight vectors $\bm w_{max}$ and $\bm w_{avg}$ are summed and then channel-wise multiplied with the input spectral feature $\bm {T}_{train}$. This process yields the final spectral feature map, denoted as $\tilde {\bm T}_{train}$, with dimensions $1 \times 24 \times 64$.  The calculation expression for the $i$-th channel feature map, denoted as $\tilde{\bm{t}}_{\text{train}_i}$ is as follows:\par
\begin{equation}\label{f1}
\begin{aligned}
\bm {w}_{total} = \bm {w}_{max}+\bm {w}_{avg}
\end{aligned}
\end{equation}
\begin{equation}\label{f1}
\begin{aligned}
\bm {\tilde{t}}_{train_i} = \bm{F}_{scale}( w_{total_i}, \bm {t}_{train_i}) = w_{total_i}\bm {t}_{train_i}
\end{aligned}
\end{equation}
where $\bm{t}_{train_i}$ ($i=1,2,\cdots, C$, and $C=C_3=64$) denotes the $i$-th channel feature map of the input feature map $\bm{T}_{train}$. $\bm{w}_{total}$ denotes the sum of the channel-wise weights obtained through maximum pooling and average pooling, with $w_{total_i}$ indicating the weight value assigned to the $i$-th channel, which is a scalar. The generated feature map is then fed into a fully connected neural network for classification. The first layer contains 24$\times$64 neurons, corresponding to the dimensionality of the output spectral signal feature $\tilde{\bm{T}}_{train}$. The second layer contains 2 neurons, matching the number of categories. The learning rate is set to 0.0002, and the cross-entropy loss function is employed throughout the implementation.\par
 
\subsection{Experimental results with the channel-wise attention-based PLS-1D-CNN model for screening infected individuals and conducting comparative experiments and analysis }
\subsubsection{Experiment and results}To validate the ability of the introduced channel-wise attention-based PLS-1D-CNN model to accurately distinguish ATR-FTIR spectral signals from positive and negative samples, especially with limited sample sizes.  We employed the proposed channel-wise attention-based PLS-1D-CNN model to classify the spectral signals of nasopharyngeal swabs collected in collaboration with Beijing Youan Hospital into negative and positive categories.   A  five-fold cross-validation approach are employed during the model implementation. This involved using 80\% of spectral signal dataset for training and reserving 20\% for evaluating the the model's performance. 80\% of the spectral signals, after baseline correction using the airPLS algorithm as depicted in Fig. \ref{airpls}, are directly input into the channel-wise PLS-1D-CNN model as the training set. The model outputs the corresponding true negative or positive label for training purposes.  \par
\begin{figure}[htbp]
	\centering
	\subfigure[]{ 
		\begin{minipage}[t]{0.5\linewidth}
			\centering
			\includegraphics[width=4.4cm]{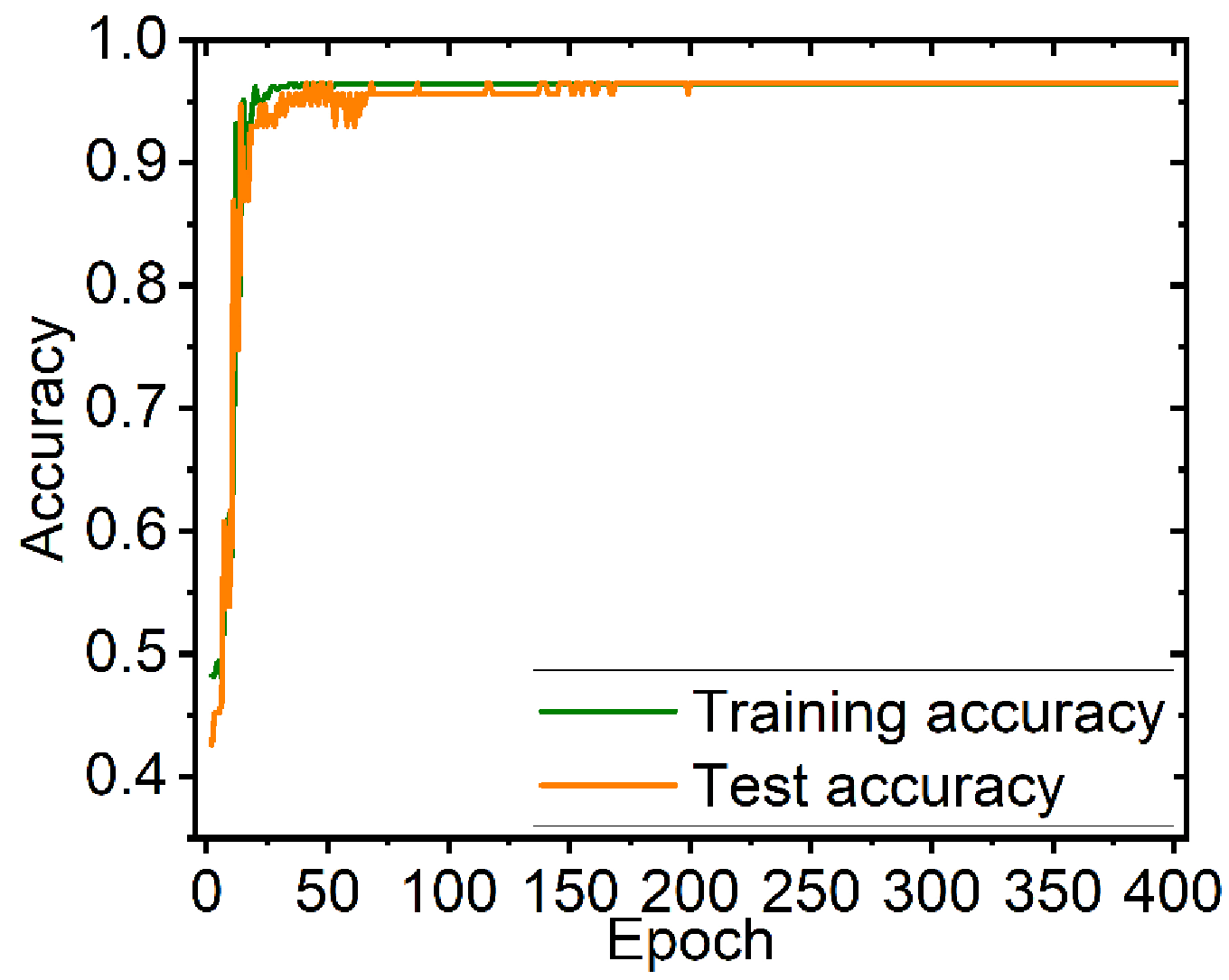}
		\end{minipage}%
	}%
	\subfigure[]{ 
		\begin{minipage}[t]{0.5\linewidth}
			\centering
			\includegraphics[width=4.4cm]{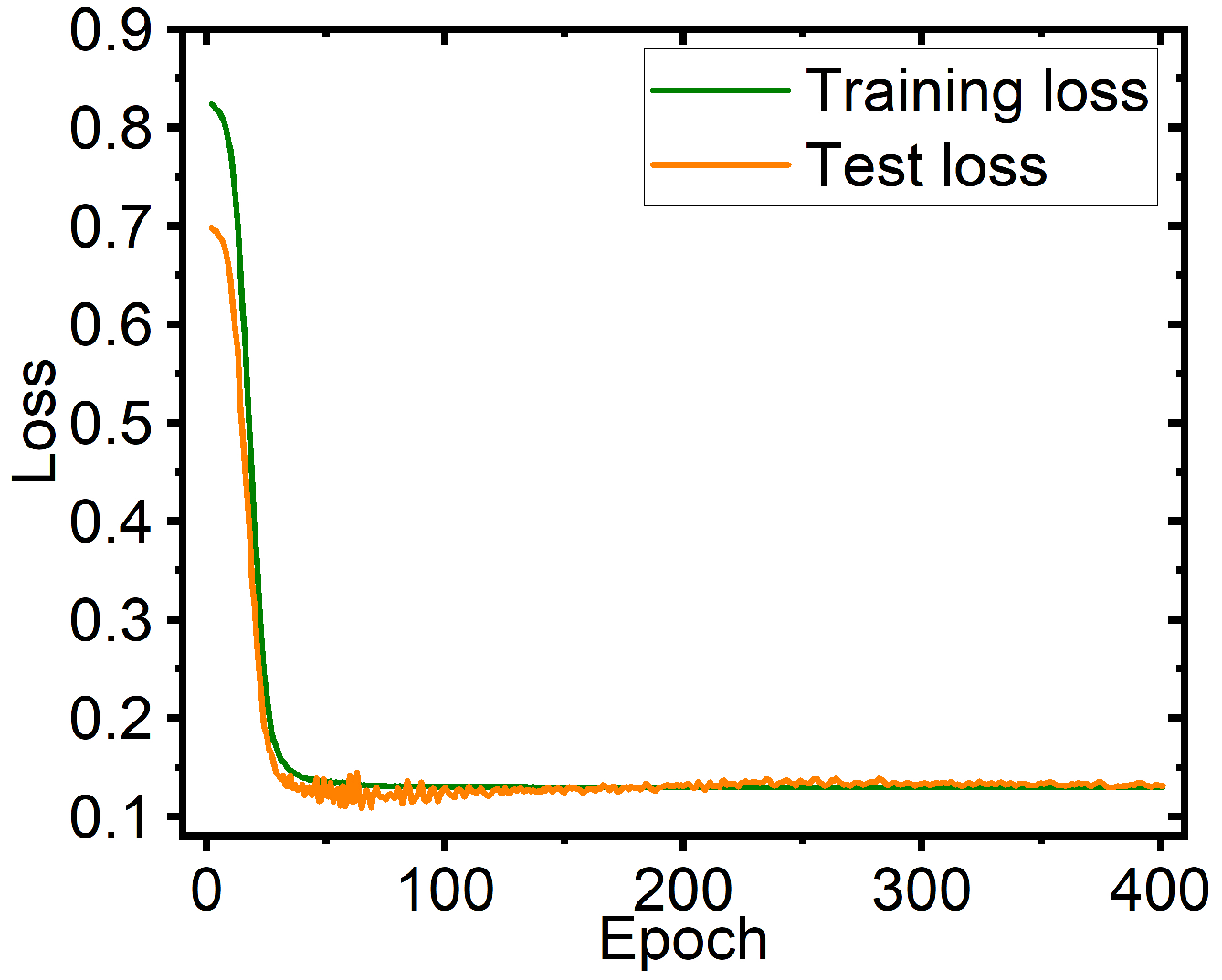}
		\end{minipage}%
	}%
	\caption{(a) The accuracy curves of the training and test sets vary as the number of training epochs changes. (b) The loss curves of the training and test sets vary as the number of training epochs changes. }
	\label{acc_loss}
\end{figure} 
\begin{figure*}[htbp]
	\centering
	\subfigure[]{ 
		\begin{minipage}[t]{0.20\linewidth}
			\centering
			\includegraphics[width=3.52cm]{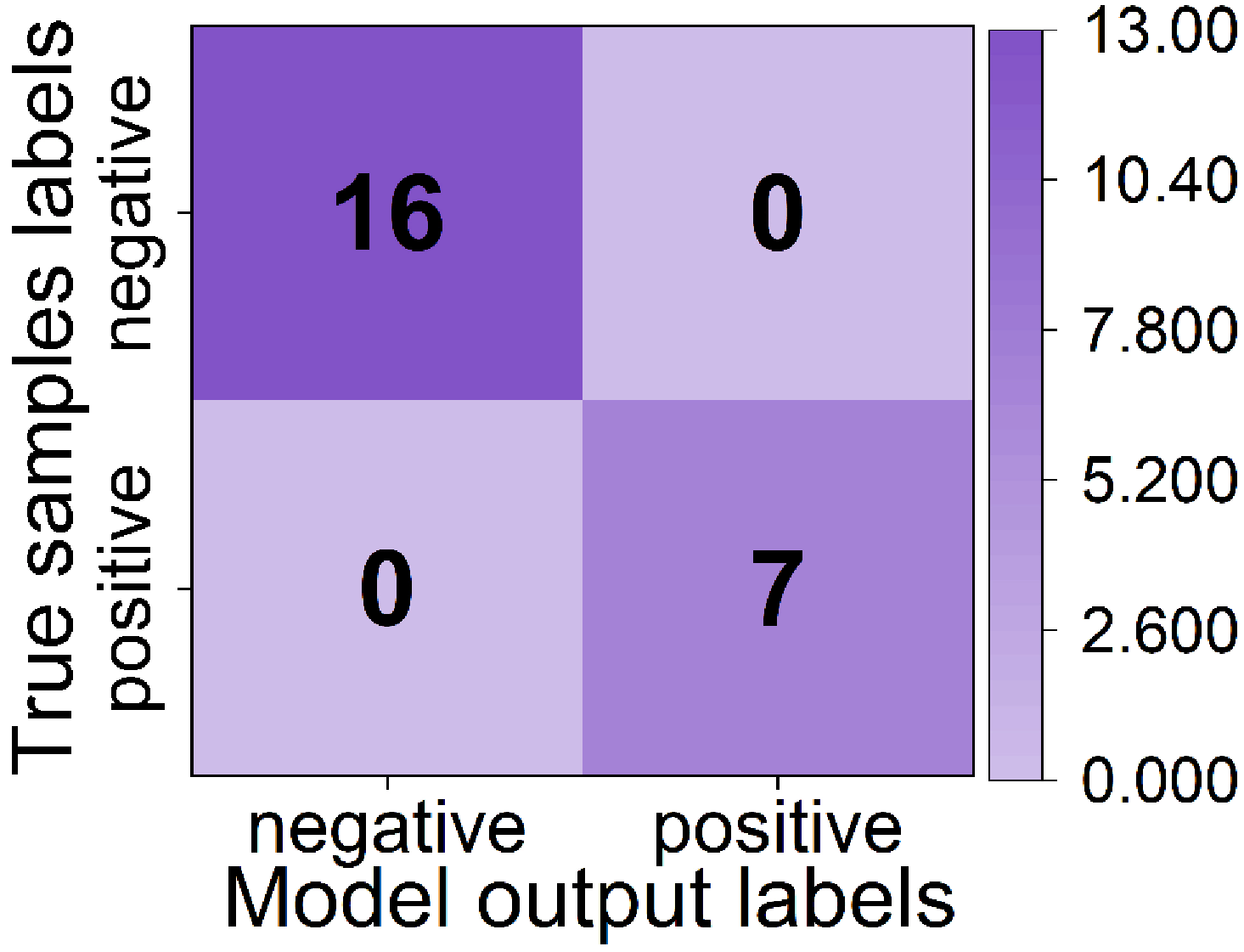}
		\end{minipage}%
	}%
	\subfigure[]{ 
		\begin{minipage}[t]{0.20\linewidth}
			\centering
			\includegraphics[width=3.52cm]{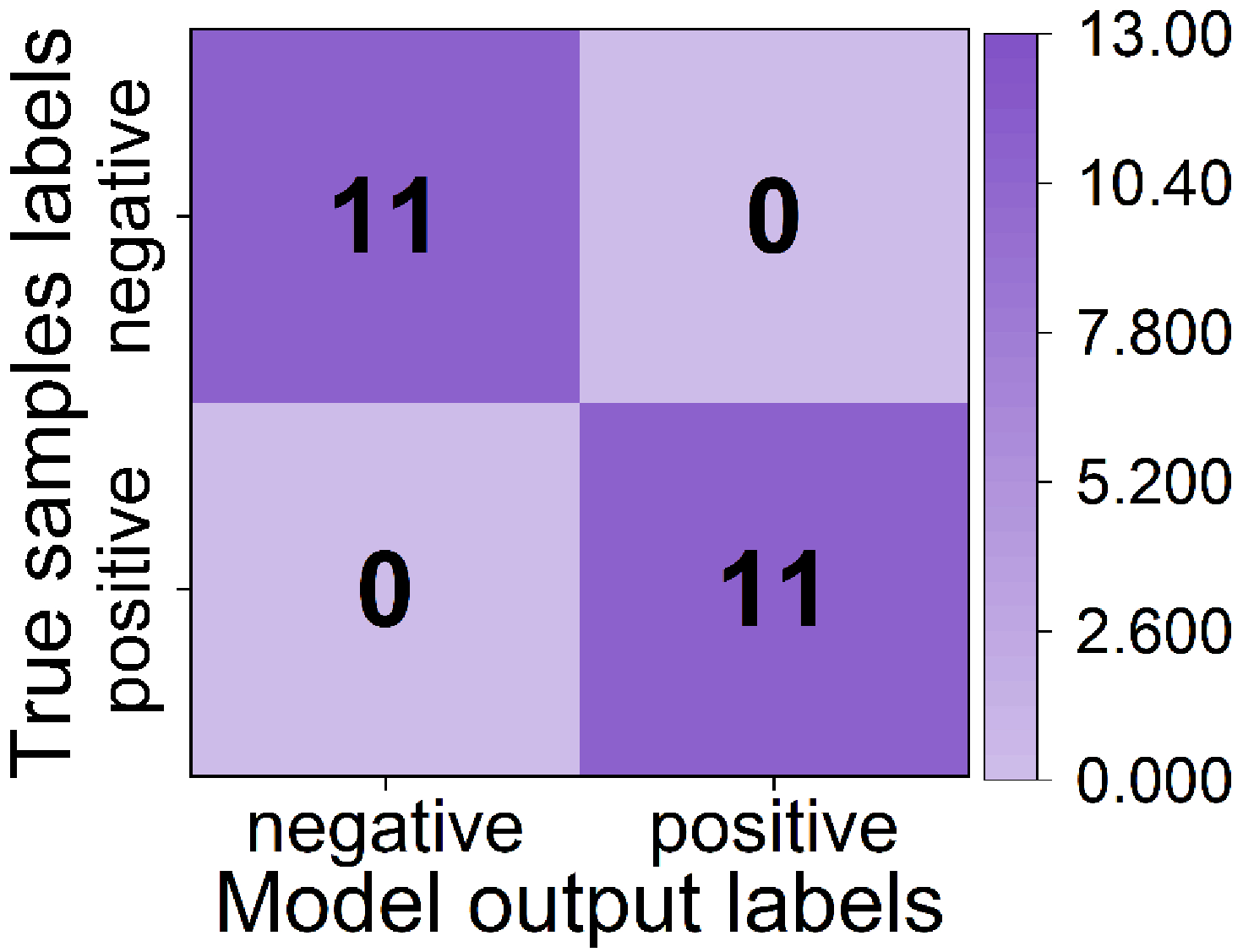}
		\end{minipage}%
	}%
	\subfigure[]{ 
		\begin{minipage}[t]{0.2\linewidth}
			\centering
			\includegraphics[width=3.52cm]{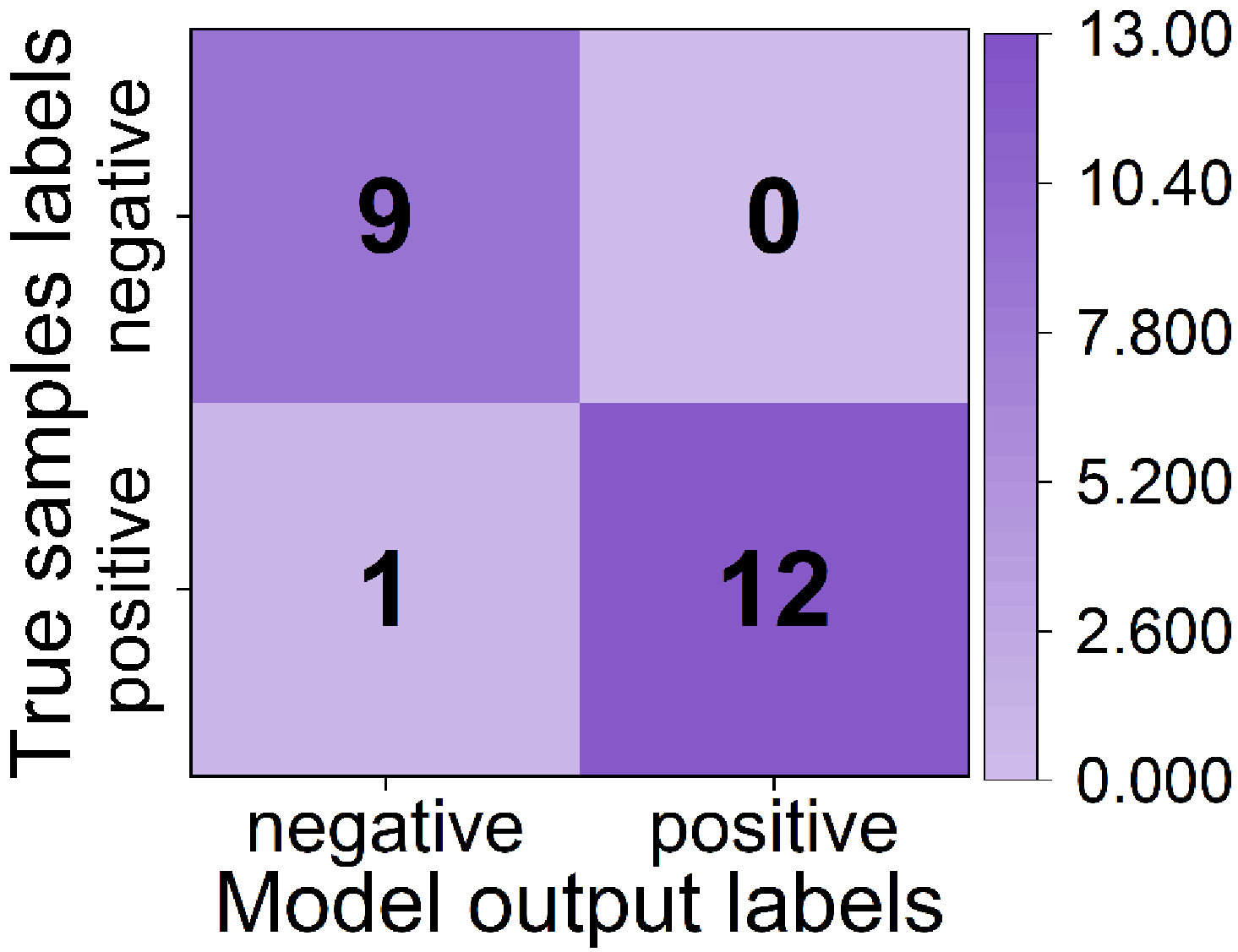}
		\end{minipage}%
	}%
	\subfigure[]{ 
		\begin{minipage}[t]{0.20\linewidth}
			\centering
			\includegraphics[width=3.52cm]{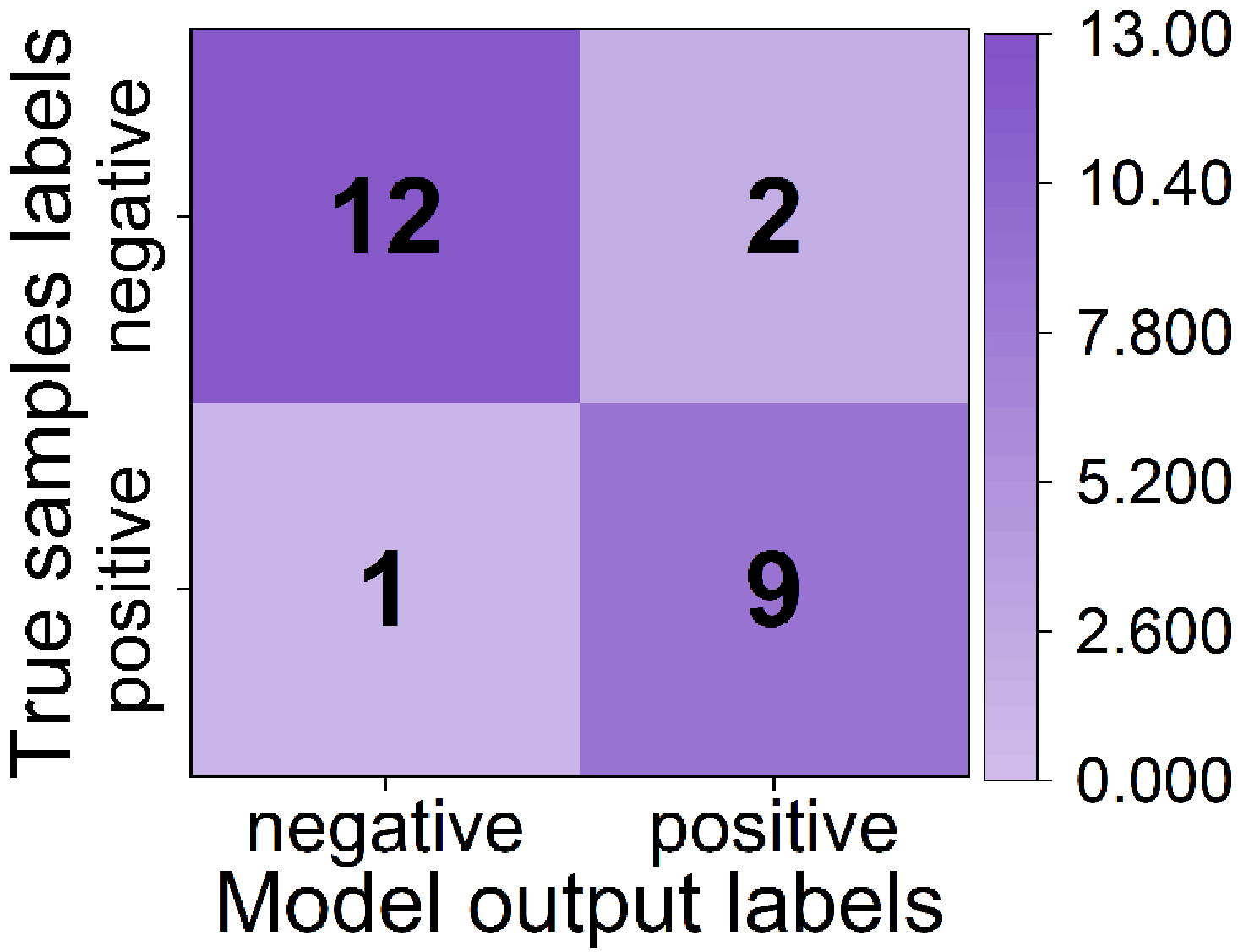}
		\end{minipage}%
	}%
	\subfigure[]{ 
		\begin{minipage}[t]{0.2\linewidth}
			\centering
			\includegraphics[width=3.52cm]{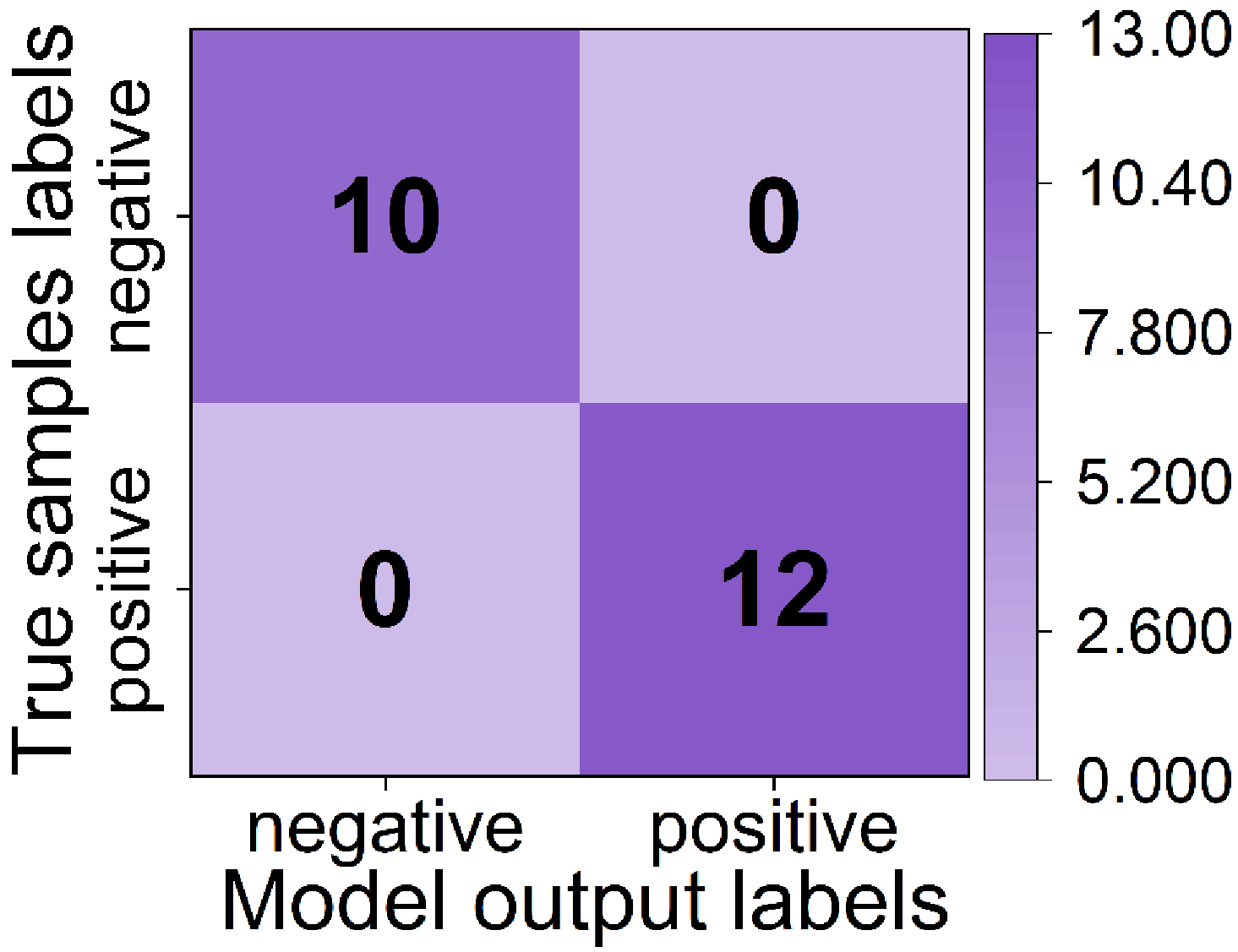}
		\end{minipage}%
	}%
	\caption{Confusion matrix results for each fold of the test set in the five-fold cross-validation. (a) First fold. (b) Second fold. (c) Third fold. (d) Fourth fold. (e) Fifth fold. }
	\label{confusion_matrix}
\end{figure*}
 
\begin{figure*}[htbp]
	\centering
		\subfigure[]{ 
		\begin{minipage}[t]{0.2\linewidth}
			\centering
			\includegraphics[width=3.5cm,height=2.8cm]{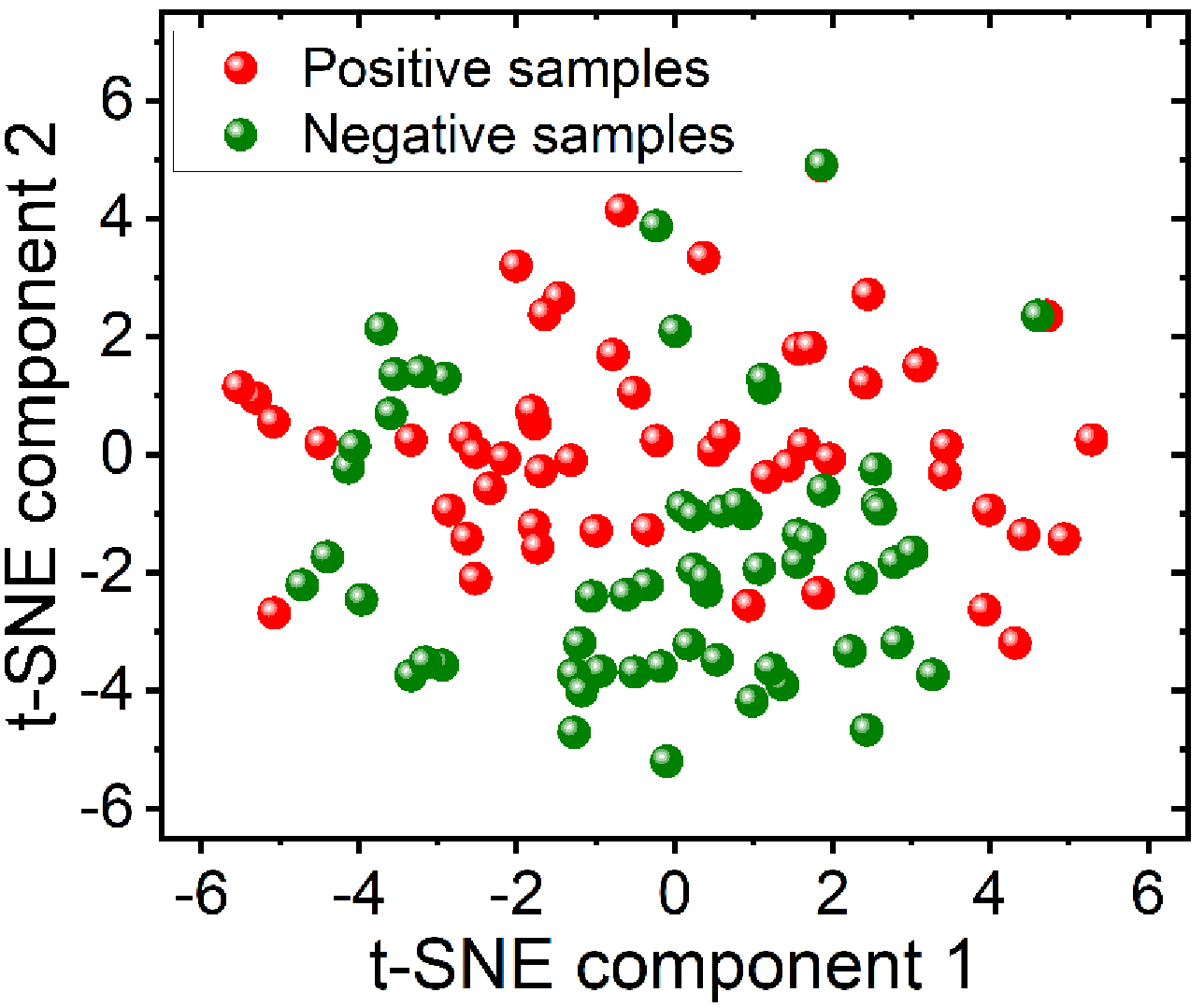}
		\end{minipage}%
	}%
	\subfigure[]{ 
		\begin{minipage}[t]{0.2\linewidth}
			\centering
			\includegraphics[width=3.5cm]{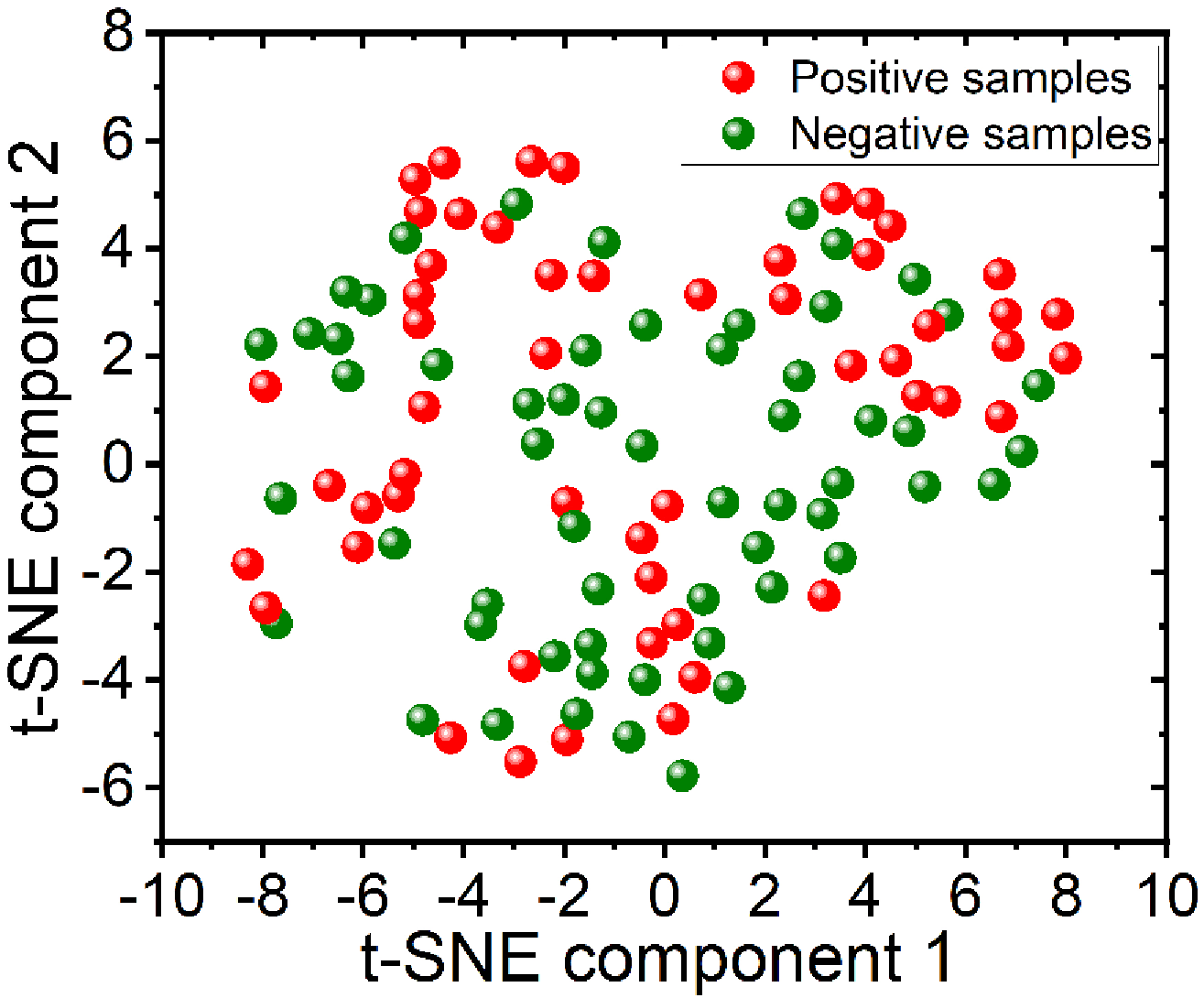}
		\end{minipage}%
	}%
	\subfigure[]{ 
		\begin{minipage}[t]{0.2\linewidth}
			\centering
			\includegraphics[width=3.5cm, height=2.8cm]{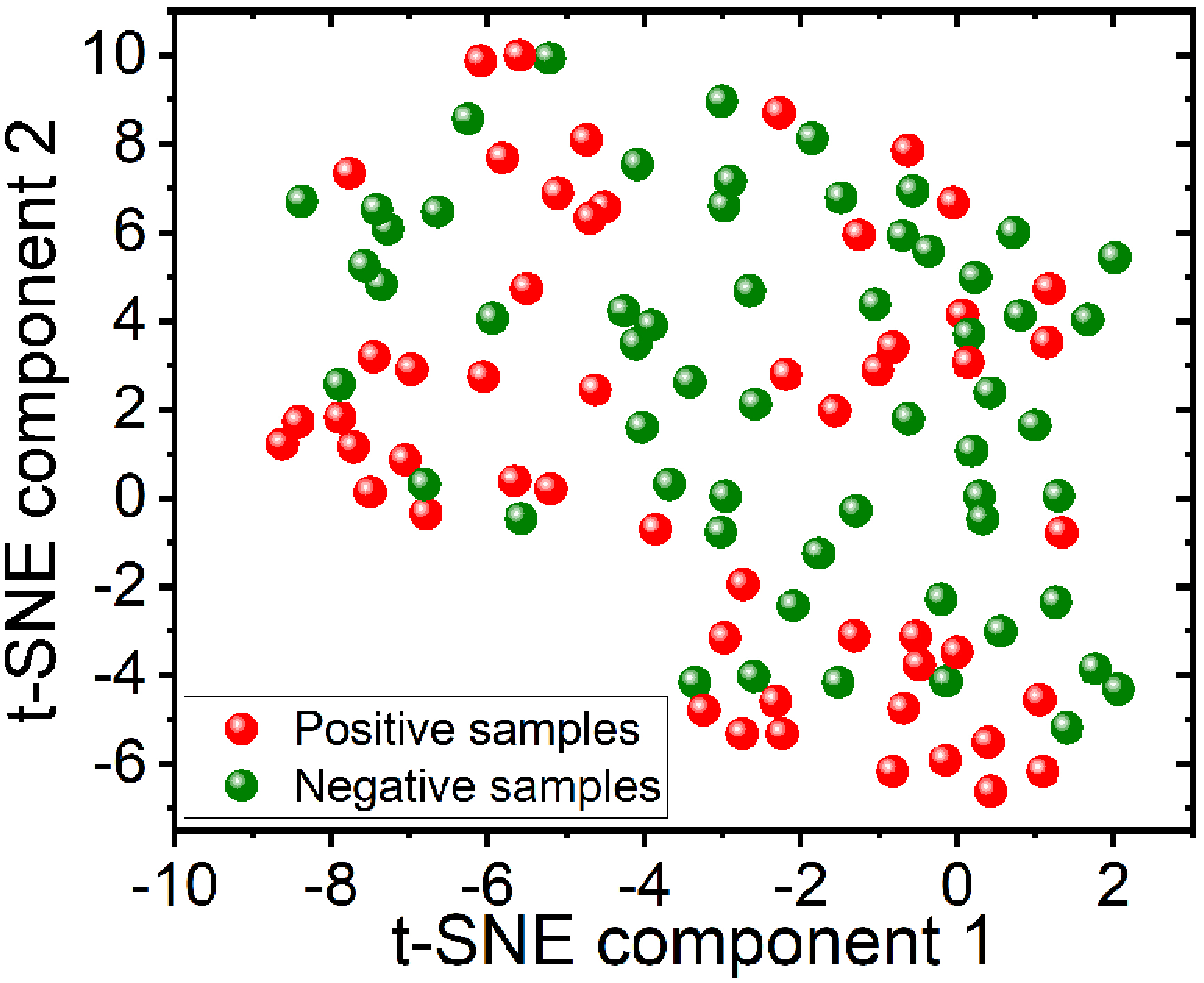}
		\end{minipage}%
	}%
	\subfigure[]{ 
		\begin{minipage}[t]{0.2\linewidth}
			\centering
			\includegraphics[width=3.5cm,height=2.8cm]{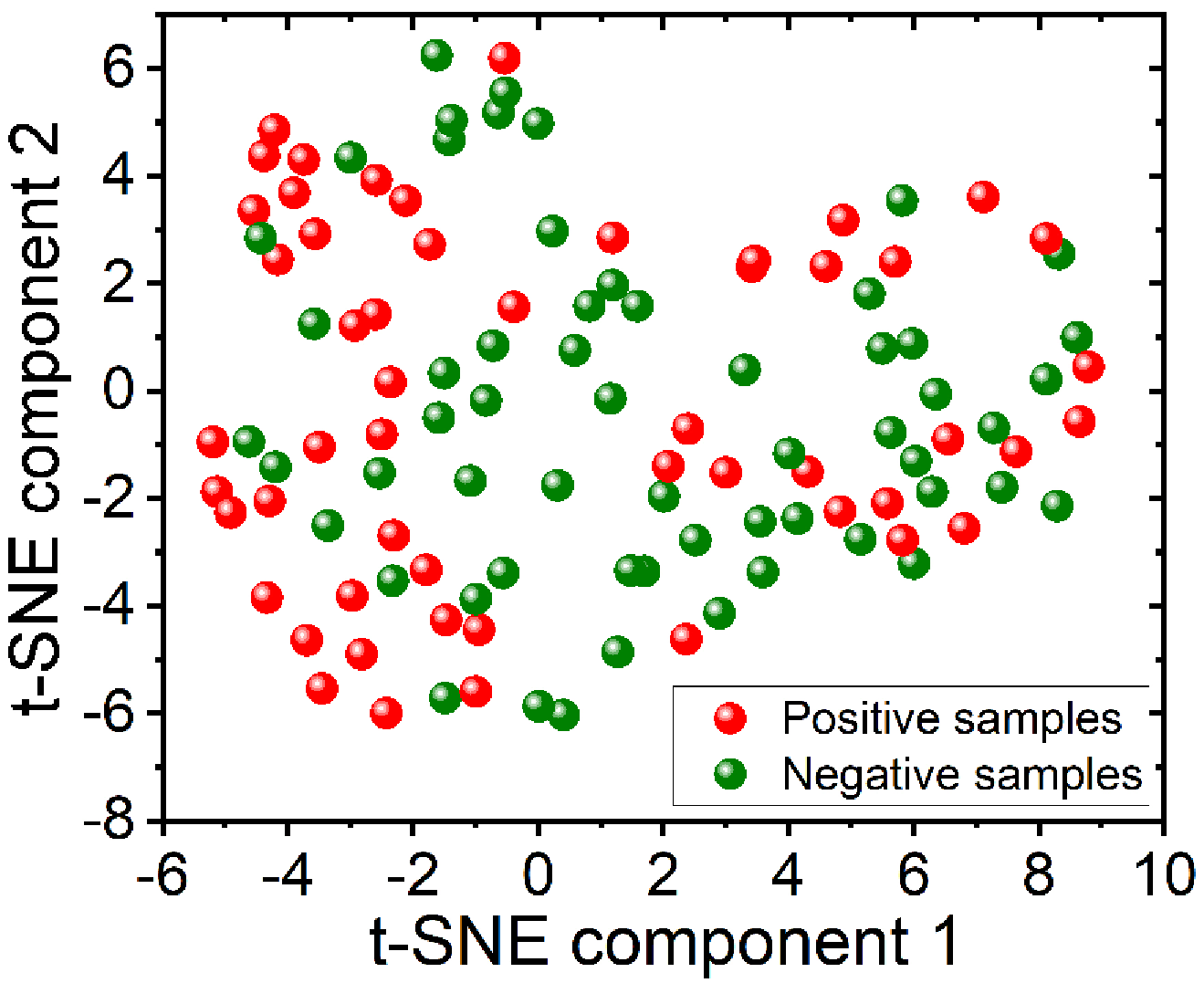}
		\end{minipage}%
	}%
	\subfigure[]{ 
		\begin{minipage}[t]{0.2\linewidth}
			\centering
			\includegraphics[width=3.5cm,height=2.8cm]{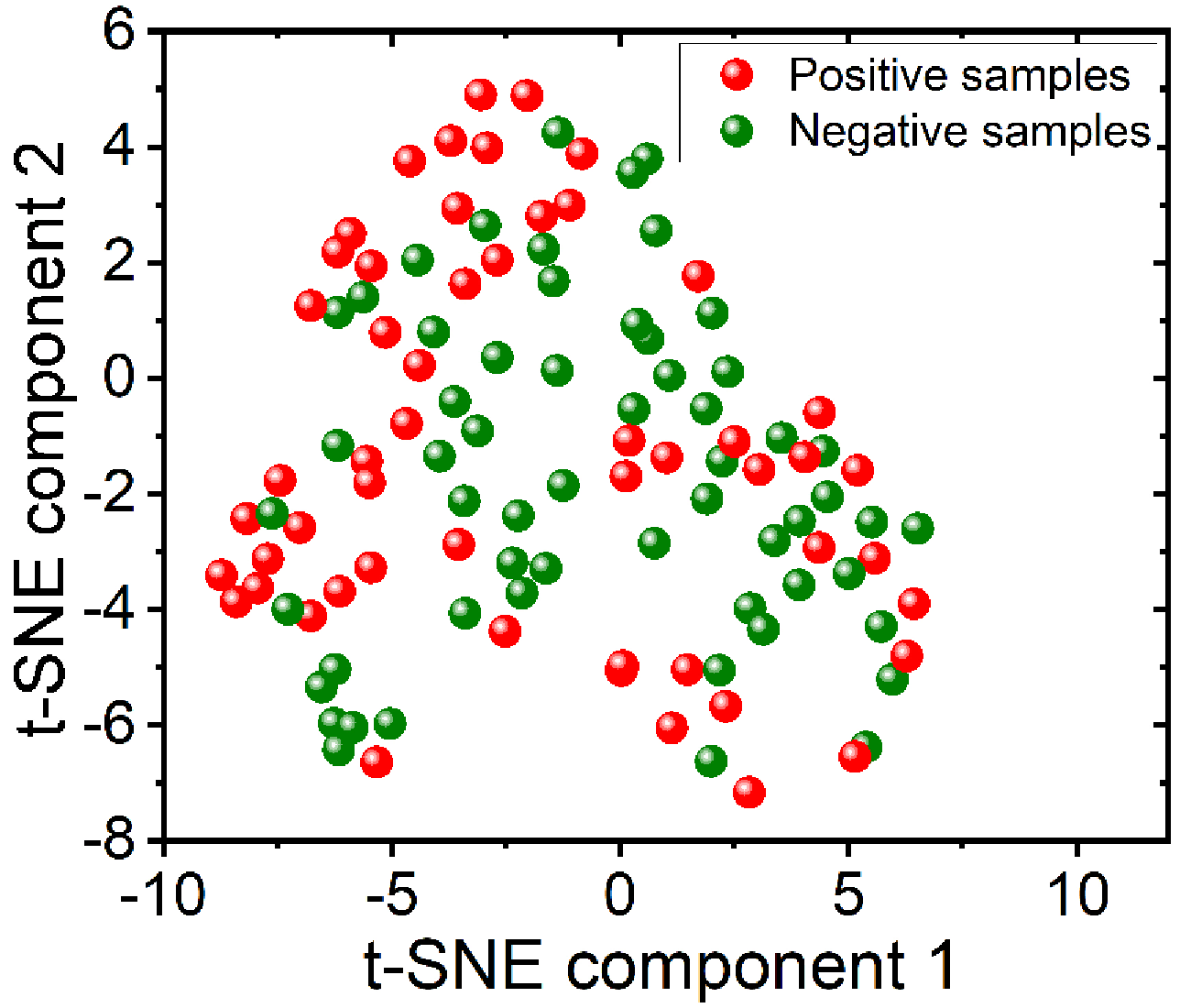}
		\end{minipage}%
	}%
	\caption{Dimensionality reduction of spectral features extracted by various feature extractors.  (a) PLS submodel feature extractor. (b) SPA feature extractor. (c) MRMR feature extractor. (d) reliefF feature extractor. (e) chi-square test feature extractor. }
	\label{tsne}
\end{figure*}

\begin{figure*}[htbp]
	\centering
	\subfigure[]{ 
		\begin{minipage}[t]{0.25\linewidth}
			\centering
			\includegraphics[width=4.4cm]{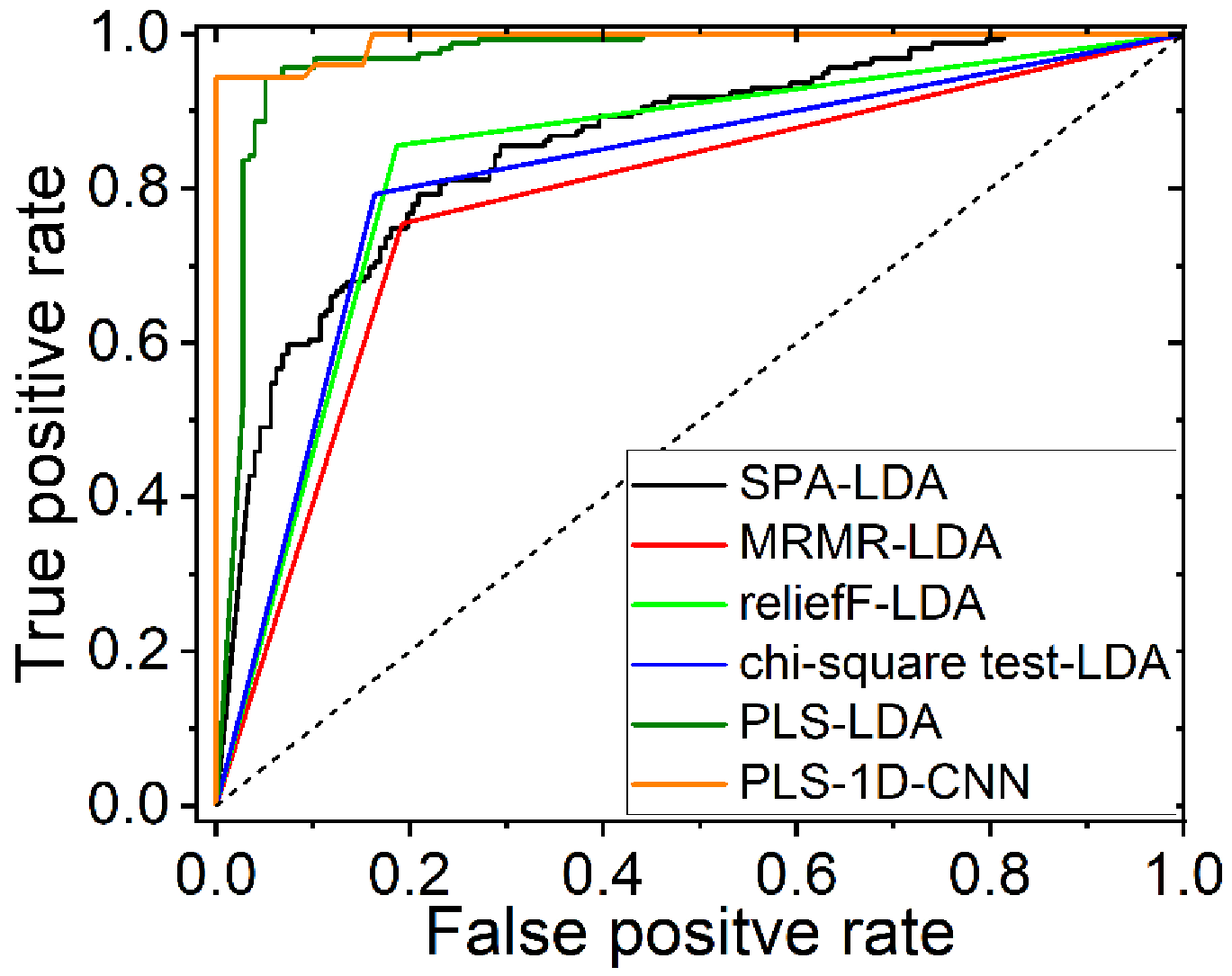}
		\end{minipage}%
	}%
	\subfigure[]{ 
		\begin{minipage}[t]{0.25\linewidth}
			\centering
			\includegraphics[width=4.4cm]{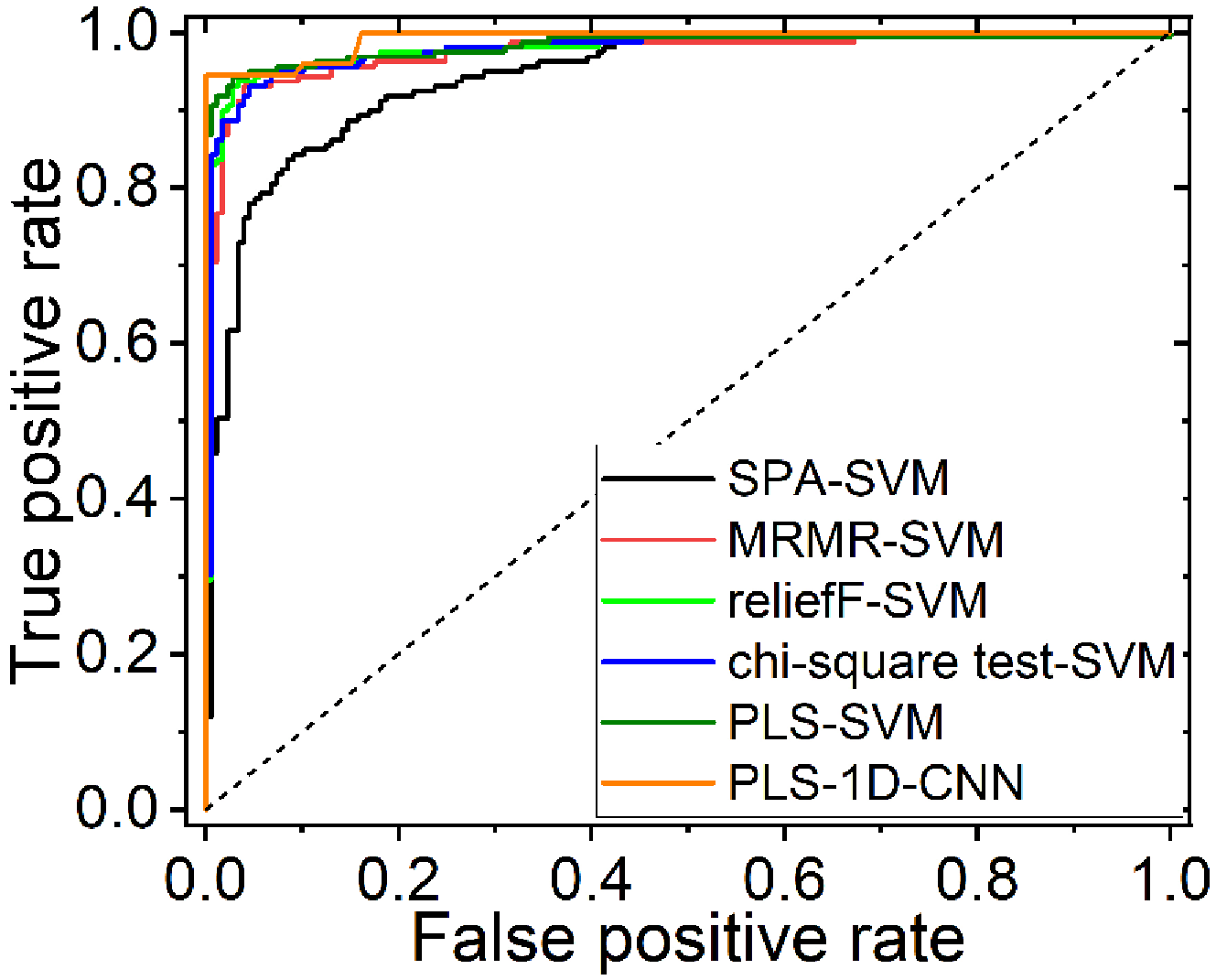}
		\end{minipage}%
	}%
	\subfigure[]{ 
		\begin{minipage}[t]{0.25\linewidth}
			\centering
			\includegraphics[width=4.4cm]{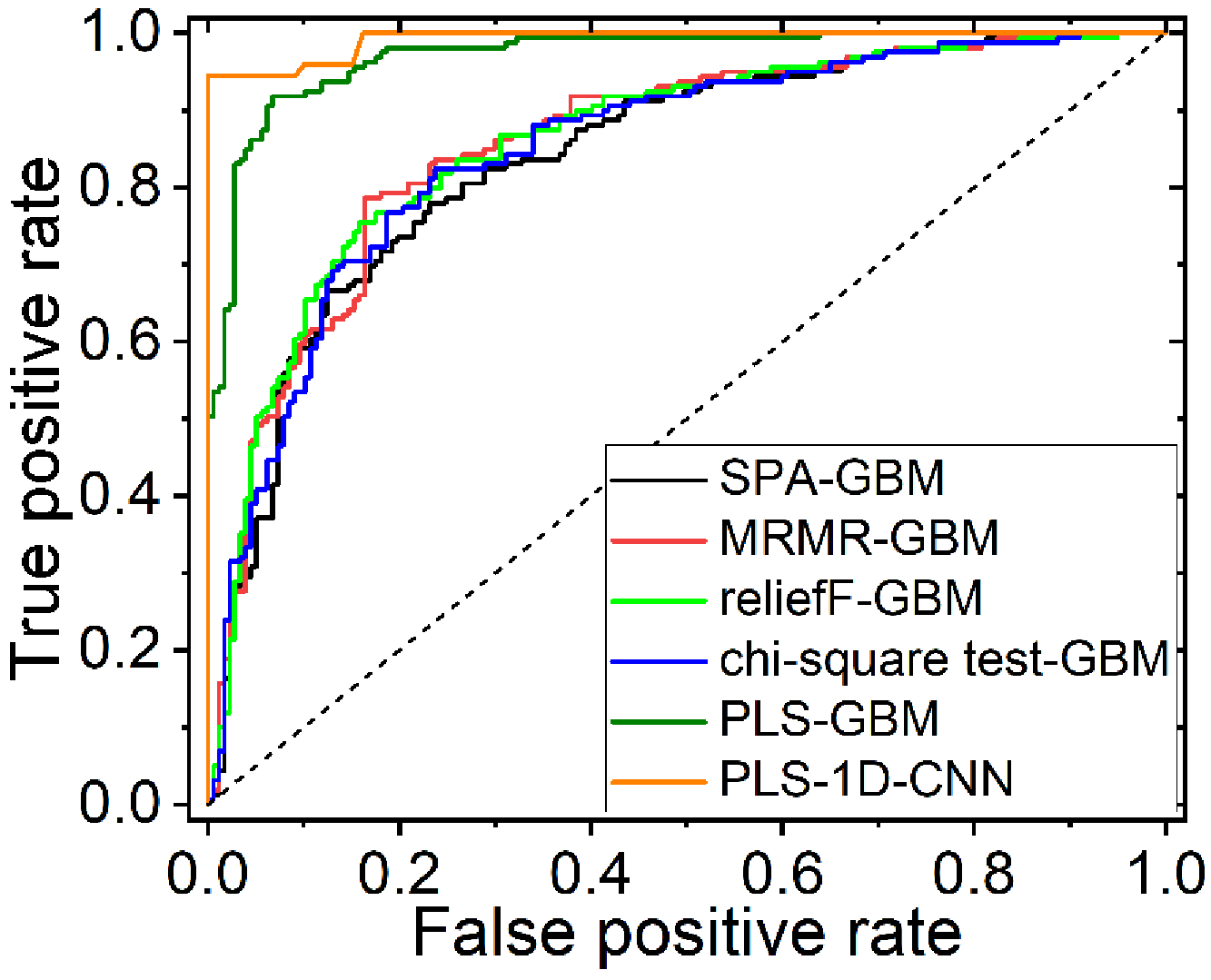}
		\end{minipage}%
	}%
	\subfigure[]{ 
		\begin{minipage}[t]{0.25\linewidth}
			\centering
			\includegraphics[width=4.4cm]{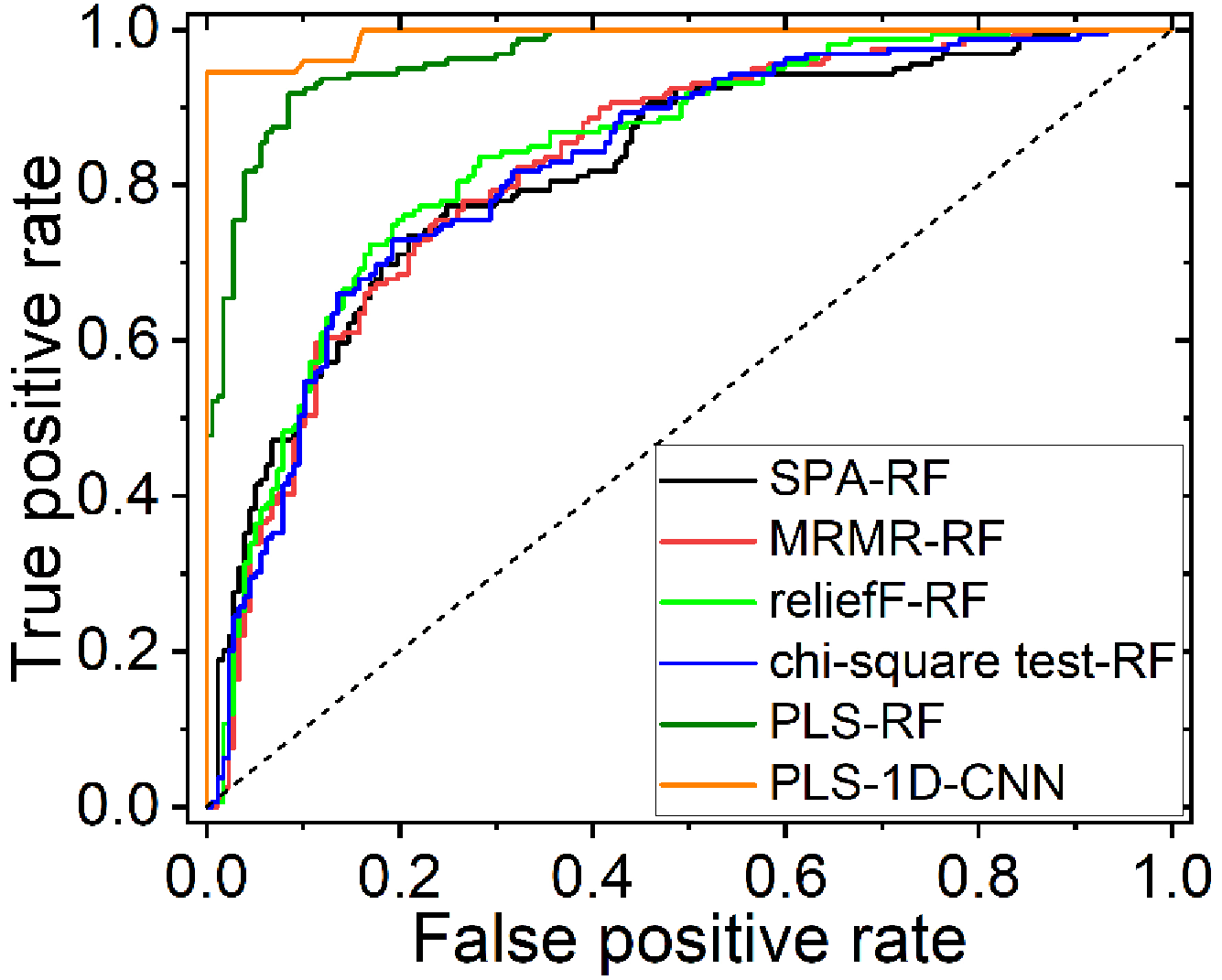}
		\end{minipage}%
	}%
	\caption{ The comparison of ROC curves for the channel-wise attention-based PLS-1D-CNN model and various classifiers combined with feature extractors, including SPA, MRMR, reliefF, chi-square test, and PLS. (a) LDA classifier. (b) SVM classifier. (c) GBM classifier. (d) RF classifier. }
	\label{ROC}
\end{figure*}  

\begin{figure}[htbp]\centering
	\includegraphics[width=6.8cm]{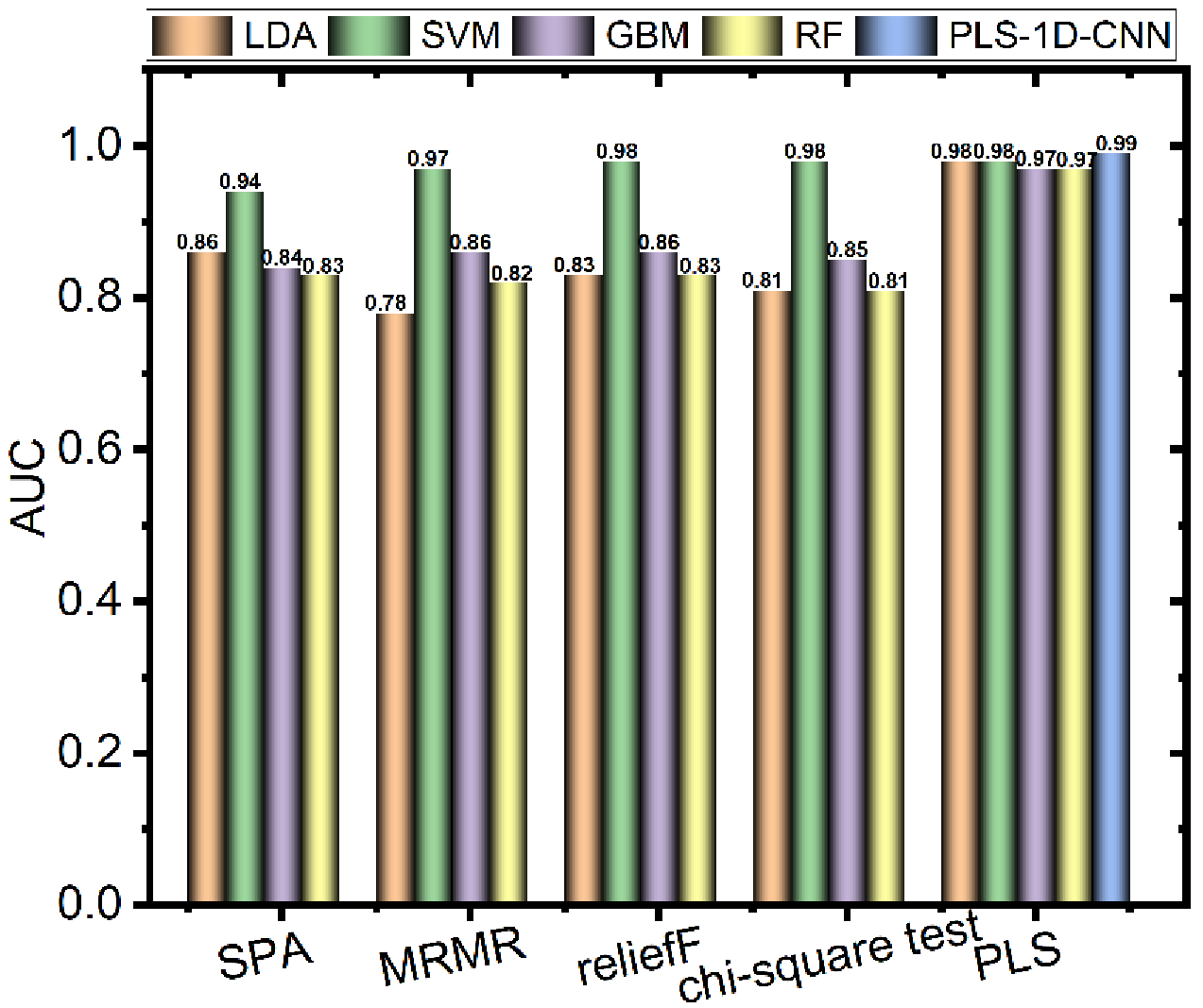}
	\caption{Comparison of AUC metrics across all evaluated models.}
	\label{AUC}
\end{figure}
        
To evaluate the model's performance more reliably, we utilize the mean accuracy and loss curves for the training and test sets obtained from five-fold cross-validation. These curves, that illustrate changes over the number of training steps, are shown in Fig. \ref{acc_loss} (a) and (b), respectively. We observe that both the accuracy curves of the training set and the test set stabilize after approximately 70 epochs, approaching close to 96.48\% without discernible differences. Likewise, in the loss function curve, the decreasing trends of the training set and test set losses are smooth and converge to the minimum stable value without notable distinctions. This indicates that the model exhibits good convergence and is not overfitting.  The results from the 5-fold cross-validation indicate that the proposed model achieves nearly 96.48\% recognition accuracy with good convergence for identifying spectral signals from nasopharyngeal swabs, especially in scenarios with limited sample sizes. During the five-fold cross-validation, each fold serves as a test set to evaluate the model. The resulting confusion matrices are presented in Fig. \ref{confusion_matrix} (a), (b), (c), (d), and (e). In the third fold test, the model misclassified one positive sample as negative, while in the fourth fold test, it misclassified two negative samples as positive and one positive sample as negative. \par 
\subsubsection{Comparative experiments}To verify the superior performance of the channel-wise attention-based PLS-1D-CNN model proposed in this study for distinguishing the infrared spectral signals of nasopharyngeal swabs from patients infected with SARS-CoV-2 and those not infected, we compared our model with various respiratory virus spectrum classification methods reported in the past three years. These include the support vector machine (SVM) classifier \cite{b13, b30, c2} in 2021, 2022, and 2023; linear discriminant analysis (LDA) \cite{b33, b39, b43} in 2022; random forest (RF) \cite{c2, b30} in 2024; and gradient boosting machine (GBM) \cite{b28} in 2022. Additionally, we combined them with different signal feature extractors, such as the successive projections algorithm (SPA) \cite{c3}, minimum redundancy maximum relevance (MRMR) \cite{c4}, relief feature selection (reliefF) \cite{c5}, chi-square test \cite{c6}, and PLS \cite{b33, b39, b43}. The metrics of accuracy, sensitivity, specificity, and F1-score were chosen to evaluate the performance of model.\par 
After correcting the baseline of the acquired spectral signal from nasopharyngeal swabs using the airPLS preprocessing method, we applied five feature extraction methods: SPA, MRMR, ReliefF, chi-square test, and PLS. The distribution diagrams of the spectral features extracted by the PLS submodel, SPA, MRMR, ReliefF, and the chi-square test, reduced to two dimensions using t-SNE, are shown in Fig. \ref{tsne} (a), (b), (c), (d), and (e), respectively. The PLS submodel employed as the spectral feature extractor for cohort 2 demonstrates greater separability than the features extracted using SPA, MRMR, ReliefF, and the chi-square test reported in the literature. Furthermore, when compared with the feature map after baseline correction using the airPLS method, as depicted in Fig. \ref{tsne_airPLS} (b), the PLS submodel exhibits stronger separability between the spectral data of the positive and negative sample classes.  The features extracted by each extractor were then input into four different classification algorithms: LDA, SVM, GBM, and RF, for recognition purposes. This process generated a total of 20 groups of experiments for comparison.   All the comparison models have been extensively debugged and optimized with the best parameters. In the LDA classifier with linear discrimination type, no regularization term was employed due to the limited amount of data (``gamma" was set to 0). Since the results of SVM vary significantly across different feature extraction methods, we optimized the parameters for each feature extractor.  For the SPA feature extractor, SVM employs a polynomial kernel function with a polynomial order set to 4, and the box constraint parameter is set to 998.67. For MRMR feature extractor, SVM employs a polynomial kernel function with a polynomial order set to 3, and the box constraint parameter is set to 139.99.   For reliefF feature extractor, SVM employs a polynomial kernel function with a polynomial order set to 2, and the box constraint parameter is set to 624.44. For chi-square test feature extractor, SVM employs a polynomial kernel function with a polynomial order set to 2, and the box constraint parameter is set to 942.52.  For PLS feature extractor, SVM employs a gaussian  kernel function with a polynomial order set to 5, and the scale parameter of the gaussian kernel function is set to 10.864. The GBM model employs the AdaBoostM1 method for training, with 80 learning cycles and a learning rate set to 0.6.  The number of trees in the random forest is set to 60, the minimum number of samples per leaf node to 5, and the number of predictors considered at each split to 5. The experimental  results for accuracy, sensitivity, specificty and F1-score  are presented in Table \ref{metrics}. SPA, MRMR, reliefF, and chi-square test feature extraction methods extract features from the nasopharyngeal swab ATR-FTIR spectrum signal, yielding 200, 600, 600, 700, and 24 features, respectively.\par
To further evaluate the classification performance of our proposed channel-wise attention-based PLS-1D-CNN model and the compared classifiers, including LDA, SVM, GBM, and RF, combined with different feature extractors such as SPA, MRMR, ReliefF, chi-square test, and PLS, we recorded the receiver operating characteristic (ROC) curves for each model.  The ROC curves for the channel-wise attention-based PLS-1D-CNN model and various classifiers (LDA, SVM, GBM, and RF) combined with feature extractors (SPA, MRMR, reliefF, chi-square test, and PLS) are depicted in Fig. \ref{ROC} (a), (b), (c), and (d), respectively. The area under the ROC curve (AUC) for each model, shown in Fig. \ref{AUC}, denotes the area under the ROC curve for each respective model. Our proposed channel-wise attention-based 1D-CNN model outperforms all other models, achieving the highest performance with accuracy of 96.48\%, sensitivity of 96.24\%, specificity of 97.14\%, and F1-score of 96.12\%, along with an AUC score  to 0.99. For the LDA, GBM, and RF classifiers, the recognition accuracy, specificity, and sensitivity are significantly higher when using PLS as the feature extractor compared to other feature extractors. Similarly, as depicted in Fig. \ref{AUC}, the AUC for the PLS feature extractor is significantly larger than that of the other feature extractors. In particular, PLS-SVM achieved a high recognition accuracy of 94.97\%, with a sensitivity of 94.70\% and a specificity of 95.89\%.  For classifiers using the same feature extractor, SVM achieved higher recognition accuracy, specificity, and sensitivity compared to the other classifiers. When utilizing the reliefF feature extractors, SVM classifiers exhibited notable performance with a high accuracy of 94.69\%, a sensitivity of 92.83\%, and a specificity of 95.89\%. Likewise, in the AUC curve, employing these feature extractors resulted in a significantly larger AUC for the SVM classifier compared to other classifiers. For LDA classifier, the recognition accuracy, sensitivity, and specificity of different feature extractors vary significantly. The accuracy, sensitivity, and specificity of the PLS feature extractor are 16.19\%, 19.34\%, and 12.00\% higher than those of the SPA feature extractor, respectively. \par
\begin{table*}[htbp]
	\centering
	\caption{Comparison results of the proposed channel-wise attention-based PLS-1D-CNN model with various classifiers combined with different feature extraction methods.}
	\begin{tabular}{ccc cccc}
		\toprule\toprule	
		\textbf{Model} & \textbf{\begin{tabular}[c]{@{}c@{}}Feature extractor\end{tabular}} & \textbf{Number of Features} & \textbf{Accuracy (\%)} & \textbf{Sensitivity (\%)} & \textbf{Specificity (\%)} & \textbf{F1-score (\%)} \\
		\midrule
		\multirow{5}{*}{LDA} & SPA & 200 & 77.87 & 76.04 & 80.81 & 76.20 \\
		& MRMR & 600 & 78.18 & 76.72 & 80.89 & 76.32 \\
		& reliefF & 600 & 83.35 & 86.36 & 81.24 & 82.84 \\
		& chi-square test & 700 & 81.54 & 79.58 & 83.94 & 79.76 \\
		& PLS & 24 & 94.06 &  \underline{95.38} & 92.81 & 93.31 \\
		\midrule 
		\multirow{5}{*}{SVM} & SPA & 200 & 86.93 & 78.81 & 93.58 & 84.06 \\
		& MRMR & 600 & 94.08 & 91.33 & \underline{96.56} & 92.77 \\
		& reliefF & 600 & 94.69 & 92.83 &  95.89 & 93.30 \\
		& chi-square test & 700 & 94.11 & 91.62 &  96.00 & 92.55 \\
		& PLS & 24 & \underline{94.97} & 94.70 & 95.89 &  \underline{94.32} \\
		\midrule
		\multirow{5}{*}{GBM} & SPA & 200 & 76.55 & 73.25 & 81.52 & 74.58 \\
		& MRMR & 600 & 79.84 & 75.98 & 84.83 & 77.39 \\
		& reliefF & 600 & 79.20 & 72.94 & 87.03 & 76.29 \\
		& chi-square test & 700 & 78.02 & 71.55 & 85.67 & 75.05 \\
		& PLS & 24 & 91.05 & 87.60 & 93.68 & 89.74 \\
		\midrule
		\multirow{5}{*}{RF} & SPA & 200 & 75.93 & 72.44 & 79.86 & 73.21 \\
		& MRMR & 600 & 75.35 & 72.00 & 79.15 & 72.81 \\
		& reliefF & 600 & 77.97 & 72.81 & 83.80 & 75.19 \\
		& chi-square test & 700 & 76.52 & 71.56 & 81.51 & 73.45 \\
		& PLS & 24 & 91.37 & 91.39 & 92.08 & 90.33 \\
		\midrule
		\multicolumn{2}{c}{\begin{tabular}[c]{@{}c@{}}Channel-wise attention-based \\ PLS-1D-CNN\end{tabular}} & 24 & \textbf{96.48} & \textbf{96.24} & \textbf{97.14} & \textbf{96.12} \\
		\bottomrule \bottomrule
	\end{tabular}
	\label{metrics}
\end{table*}
\begin{table*}[htbp]
	\centering  
	\caption{Comparison of the diagnostic performance of various reported methods for COVID-19}
	\scalebox{0.80}{
		\begin{tabular}{ccc cccc}
			\toprule\toprule	
			\textbf{Technique
			} & \textbf{\begin{tabular}[c]{@{}c@{}}Clinical samples\\(P: positive; N: negative)\end{tabular}} & \textbf{Detection time 
				(min)} & \textbf{Accuracy (\%)} & \textbf{Sensitivity (\%)} & \textbf{Specificity (\%)} & Reference                                                                                                         
			
			\\ \hline                      
			\begin{tabular}[c]{@{}c@{}}Nucleic acid testing
			\end{tabular}&    \begin{tabular}[c]{@{}c@{}} Oropharyngeal swab (P:371, N:352)\end{tabular}  & 45	&        97.73 & 99.73 &98.76 & \cite{s2}
			
			\\                           
			\begin{tabular}[c]{@{}c@{}}Antigen testing
			\end{tabular}&    \begin{tabular}[c]{@{}c@{}}  Nasopharyngeal swabs (P:201, N:50)\end{tabular}  & 10	&        75.6 & 100 & 80.5 & \cite{s3}                                                                                 
			
			\\                           
			\begin{tabular}[c]{@{}c@{}}Raman spectroscopy
			\end{tabular}&    \begin{tabular}[c]{@{}c@{}}  Throat swabs in VTM  (P:35, N:201)\end{tabular}  & 20	&        91.4 & 95 & 94.1  & \cite{s34}
			
			\\                           
			\begin{tabular}[c]{@{}c@{}}Infrared spectroscopy 	
			\end{tabular}&    \begin{tabular}[c]{@{}c@{}}  Saliva in VTM (P:29, N:28)\end{tabular}  & 15	&        87.72 & 93.1 & 82.14  & \cite{b44}
			
			\\                           
			\begin{tabular}[c]{@{}c@{}}Infrared spectroscopy 	
			\end{tabular}&    \begin{tabular}[c]{@{}c@{}}  Pharyngeal swabs (P:35, N:201)\end{tabular}  & 2	&        91.4 & 95 & 94.1  & \cite{c1}
			
			\\                        
			\begin{tabular}[c]{@{}c@{}}Infrared spectroscopy
			\end{tabular}&    \begin{tabular}[c]{@{}c@{}}  Nasopharyngeal swabs in VTM (P:53, N:59)\end{tabular}  & 10	&        96.24 & 97.14 & 96.48  & \begin{tabular}[c]{@{}c@{}}  This work\end{tabular}
			
			\\
			\bottomrule \bottomrule
	\end{tabular}}
	\label{comparison}
\end{table*} 
\subsubsection{ Ablation experiment}
\begin{figure}[htbp]
	\centering
	\subfigure[]{ 
		\begin{minipage}[t]{0.5\linewidth}
			\centering
			\includegraphics[width=4.3cm]{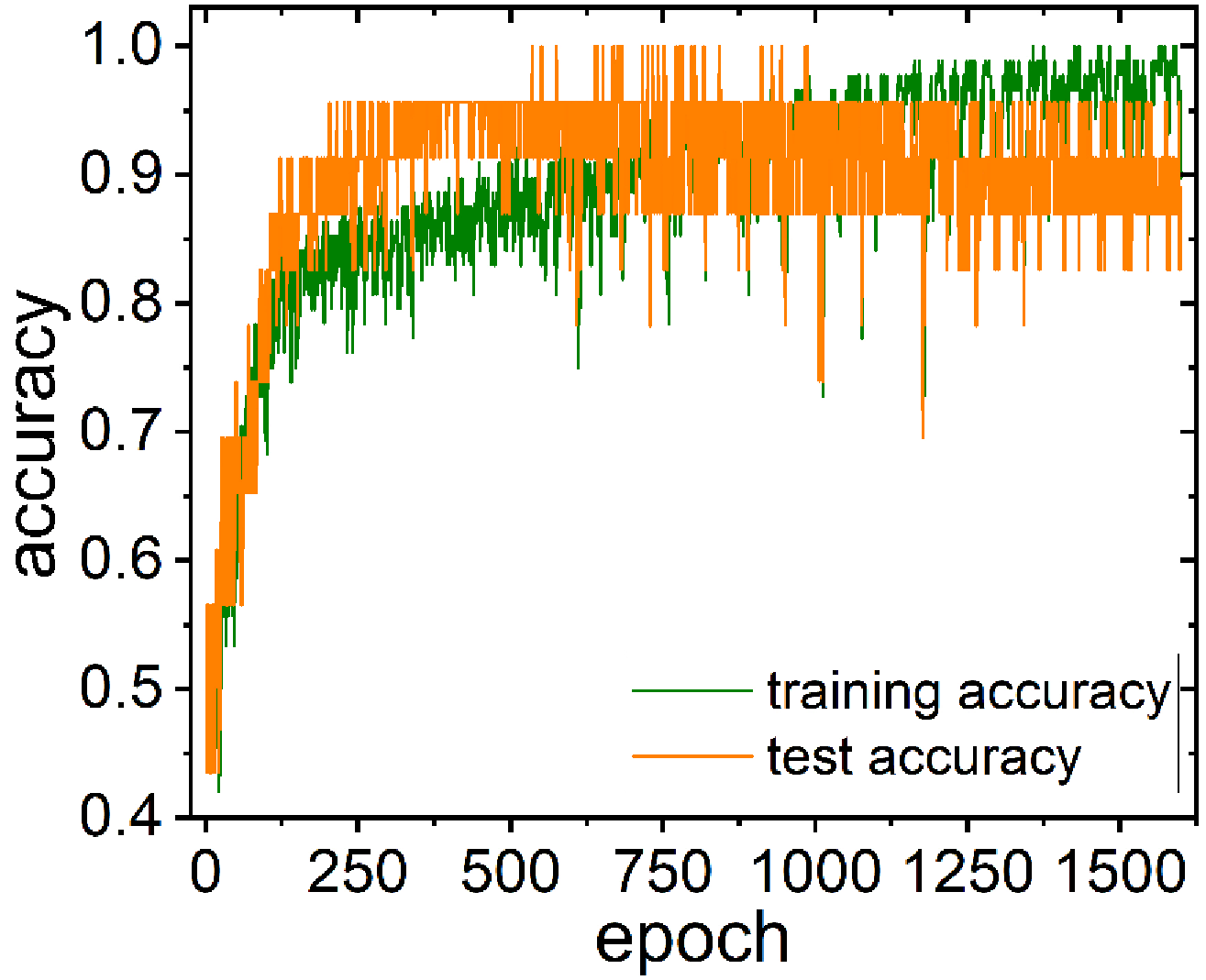}
		\end{minipage}%
	}%
	\subfigure[]{ 
		\begin{minipage}[t]{0.5\linewidth}
			\centering
			\includegraphics[width=4.4cm]{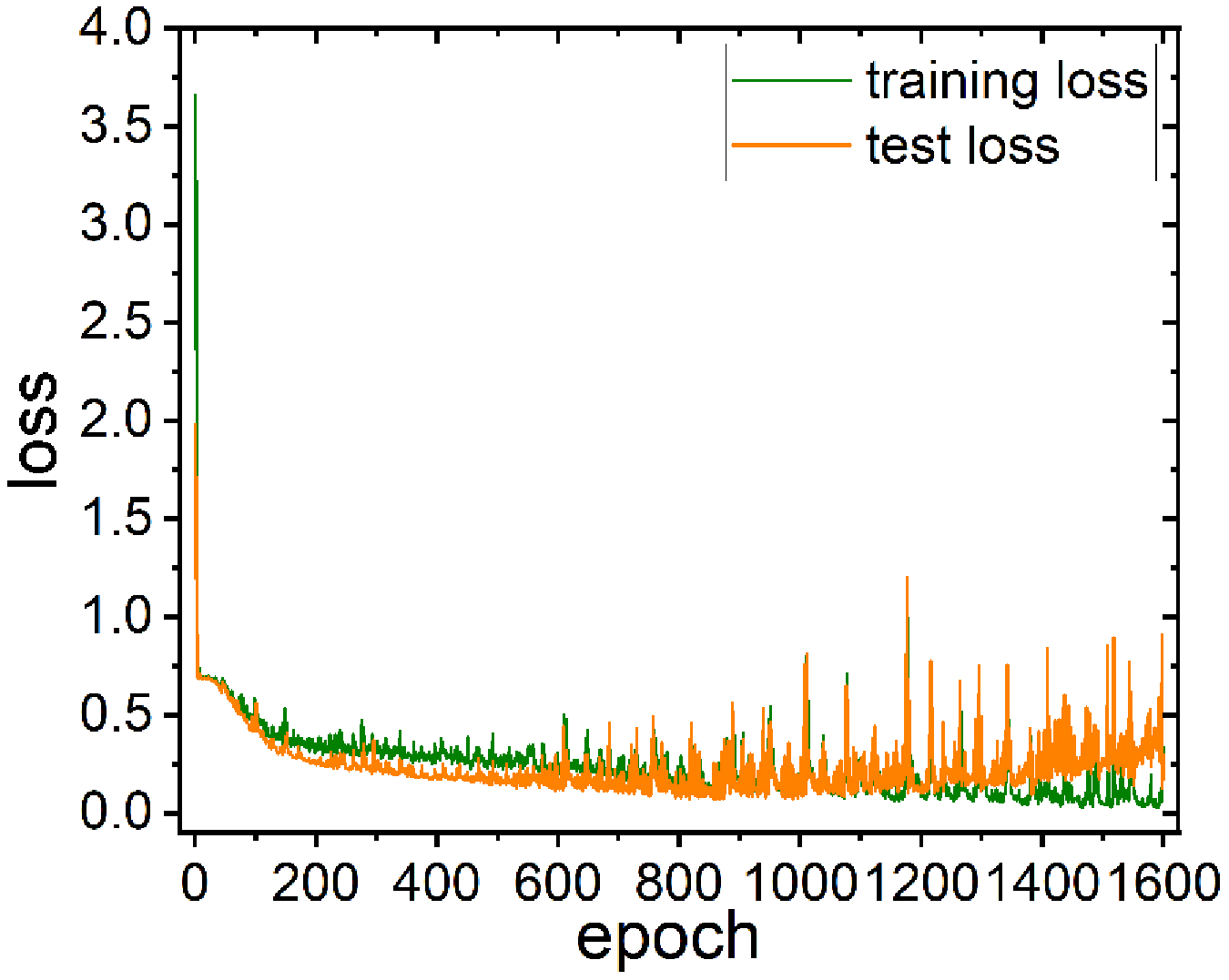}
		\end{minipage}%
	}%
	\caption{(a) The accuracy curves of the training and test sets vary as the number of training epochs changes. (b) The loss curves of the training and test sets vary as the number of training epochs changes. }
	\label{acc_loss_874}
\end{figure}

To further demonstrate that the PLS submodel can effectively reduce the dimensionality of spectral data while extracting highly informative features correlated with the target variable, enhancing the recognition ability of infrared spectral signals for distinguishing between positive and negative nasopharyngeal swabs, we conducted an ablation experiment. In this experiment, we directly utilized the 874-dimensional ATR-FTIR spectral signal, after baseline correction with the airPLS algorithm, as the input for the 1D-CNN submodel. The average accuracy and loss curves after five-fold cross-validation, as the number of steps changes, are shown in Fig. \ref{acc_loss_874} (a) and \ref{acc_loss_874} (b), respectively. Compared to the accuracy and loss curves shown in Fig. 8 (a) and (b), it is evident that the training and test set curves exhibit significant fluctuations and insufficient convergence. Before reaching approximately 900 epochs, the model shows insufficient performance on the training set but performs relatively better on the test set, suggesting an incomplete learning of patterns within the training data. However, as training progresses beyond 900 epochs, the model's performance on the training set steadily improves, eventually exceeding the accuracy on the test set, signaling the onset of overfitting. As a result, we can conclude that in scenarios involving a small number of multi-dimensional spectral signals, the 1D-CNN model alone cannot effectively extract key spectral features. It requires the integration of an effective feature extractor to enhance the model's recognition accuracy and robustness.   \par  

\subsubsection{Analysis and conclusion from experiments}In summary, while the MRMR method combined with an SVM classifier achieves a high specificity of 96.56\% and a sensitivity of 91.33\%, and the chi-square test method with an SVM classifier reaches a specificity of 96.00\% and a sensitivity of 91.62\%, both fall slightly short of the World Health Organization (WHO) recommended standards \cite{s1}, which require point-of-care tests for prior SARS-CoV-2 infection to have a sensitivity above 90.00\% and a specificity exceeding 97.00\%. In contrast, our proposed channel-wise attention-based PLS-1D-CNN model demonstrates superior performance, achieving an accuracy of 96.48\%, sensitivity of 96.24\%, specificity of 97.14\%, F1-score of 96.12\%, and an AUC score of 0.99. It not only meets the WHO's point-of-care testing standards but also satisfies the more stringent requirement of 95\% sensitivity and 97\% specificity for testing prior SARS-CoV-2 infection in moderate to high-volume scenarios. The proposed screening system  leverages spectra from a limited number of collected nasopharyngeal swab samples to directly construct a model. As shown in TABLE \ref{comparison}, while the method reported in literature \cite{c1} completes the recognition task in just 2 minutes, its detection accuracy is limited to 91.4\%. In contrast, the nucleic acid detection method described in literature \cite{s2} achieves an accuracy of 97.73\%, with a sensitivity of 99.73\% and a specificity of 98.76\%. However, this improved performance requires a longer detection time of 45 minutes. Our method demonstrates superior performance by effectively balancing the critical requirements of rapid detection with high accuracy, sensitivity, and specificity, outperforming other reported COVID-19 diagnostic approaches. This system has the potential to enable repeatable and uninterrupted large-scale non-destructive screening, near real-time on-site online measurement, and identification, while keeping inspection costs low and maintaining high accuracy, sensitivity, and specificity.\par

\section{conclusion}
In this study, we developed a system that combines ATR-FTIR with airPLS preprocessing and a channel-wise attention-based PLS-1D-CNN model to accurately identify SARS-CoV-2 infections within 10 minutes, using a limited number of nasopharyngeal swabs from suspected patients. Given that ATR-FTIR spectra are highly susceptible to variations in experimental conditions, which can affect their quality, we proposed a biomolecular importance (BMI) evaluation method to quantitatively assess the significance of virus-related biomolecules in distinguishing high-quality spectral signals collected under different conditions. This approach reveals the underlying biological correlations, facilitating the selection of superior spectra and standardizing experimental procedures to ensure consistent and reliable signal collection. Additionally, the proposed channel-wise attention-based PLS-1D-CNN model integrates the linear feature extraction capability of PLS, effectively handling multicollinearity and high correlation in spectral data by transforming features into a low-dimensional space that retains the most relevant information for classification labels, while leveraging the nonlinear feature learning ability of CNN. Furthermore, the channel-wise attention mechanism dynamically adjusts each feature channel's importance, ensuring the model focuses on the most relevant features, thereby enhancing both feature selection and classification performance.  \par
Experimental verification on nasopharyngeal swabs collected in Beijing Youan Hospital from suspected SARS-CoV-2 patients validated our model's superior performance. It outperformed various classifiers such as LDA, SVM, GBM, and RF combined with feature extractors like SPA, MRMR, reliefF, chi-square test, and PLS, as reported in recent literature. This proposed screening system shows promise for early detection of new respiratory viruses and is potentially well-suited for large-scale screening at critical locations such as airports, hospitals, schools, and train stations.\par
In our future work, we have identified two  directions to pursue. Firstly, we will collect more spectral signals from nasopharyngeal swabs of SARS-CoV-2 patients, further validating the stability and consistency of the proposed channel-wise attention-based PLS-1D-CNN model. Secondly, we will enhance the model's functionality to include the identification of specific types of SARS-CoV-2 infection within the detected positive cases.\par


\begin{thebibliography}{1}	

\bibitem{b1} M. -H. Tayarani-Najaran, ``A Novel Ensemble Machine Learning and an Evolutionary Algorithm in Modeling the COVID-19 Epidemic and Optimizing Government Policies,'' \emph{IEEE Trans. Syst., Man Cybern., Syst.}, vol. 52, no. 10, pp. 6362-6372, Oct. 2022.

\bibitem{b2} J. Cai \emph{et al.}, ``Modeling transmission of SARS-CoV-2 Omicron in 
China,'' \emph{Nat. Med.}, vol. 28, no. 7, pp. 1468-1475, May 2022.


\bibitem{r3} B. Xiao \emph{et al},, ``COVID-19 transmission dynamics underlying epidemic waves in Kenya,'' \emph{Science}, vol. 374, no. 6570, pp. 989-994, Oct. 2021.

\bibitem{r4} P. Nouvellet \emph{et al}, ``Reduction in mobility and COVID-19 transmission,'' \emph{Nat. Commun. }, vol. 12, no. 1, pp. 1-9, Feb. 2021.

\bibitem{r5} E. Davis \emph{et al},, ``Contact tracing is an imperfect tool for controlling COVID-19 transmission and relies on population adherence,'' \emph{Nat. Commun. }, vol. 12, no. 1, pp. 1-8, Sep. 2021.


\bibitem{b5} T. Wang, D. Wu, and W. Z. Zhao, ``Minimizing indirect contacts in urban pick-Up and
delivery services during COVID-19 pandemic,'' \emph{IEEE Trans. Syst., Man Cybern., Syst.}, vol. 53, no. 12, pp. 7876-7887, Dec. 2023.

\bibitem{b7} L. Wang \emph{et al}, ``Rapid and ultrasensitive electromechanical detection of ions, biomolecules and SARS-CoV-2 RNA in unamplified samples,'' \emph{Nat. Biomed. Eng.}, vol. 6, no. 3, pp. 276-285, Feb. 2022.

\bibitem{r6} H. Lin \emph{et al}, ``Ferrobotic swarms enable accessible and adaptable automated viral testing,'' \emph{Nature}, vol. 611, no. 7936, pp. 570-577, Nov. 2022.

\bibitem{b8} D. R. M. Smith \emph{et al}, ``Rapid antigen testing as a reactive response to surges in nosocomial SARS-CoV-2 outbreak risk,'' \emph{Nat. Commun.}, vol. 13, no. 1, pp. 1-10, Jan. 2022.

\bibitem{b9} C. Chen \emph{et al}, ``Rapid detection of anti-SARS-CoV-2 antibody using a selenium nanoparticle-based
lateral flow immunoassay,'' \emph{IEEE Trans. Nanobiosci.}, vol. 21, no. 1, pp. 37-43, Jan. 2022.

\bibitem{r7} F. Ghorbanizamani \emph{et al}, ``Dye-loaded polymersome-based lateral flow assay: rational design of a COVID-19 testing platform by repurposing SARS-CoV-2 antibody cocktail and antigens obtained from positive human samples,'' \emph{ACS Sens.}, vol. 6, no. 8, pp. 2988-2997, Jul. 2021.

\bibitem{b10} P. Singh and Y.-P Huang ``AKDC: Ambiguous Kernel Distance Clustering Algorithm for COVID-19 CT Scans Analysis,'' \emph{IEEE Trans. Syst., Man Cybern., Syst.}, vol. 54, no. 10, pp. 6218-6229, Jul. 2024.



\bibitem{b13} A. Dairi, F. Harrou, and Y. Sun, ``Deep generative learning-based 1-SVM detectors for unsupervised COVID-19 infection
detection using blood tests,'' \emph{IEEE Trans. Instrum. Meas.}, vol. 71,  pp. 1-14, Nov. 2021.

\bibitem{b14} K. Roy \emph{et al}, ``LwMLA-NET: A Lightweight Multi-Level
Attention-Based NETwork for Segmentation of COVID-19 Lungs Abnormalities From CT Images,'' \emph{IEEE Trans. Instrum. Meas.}, vol. 71,  pp. 1-13, Mar. 2022.

\bibitem{b15} L. L. Zeng, K. Gao, D. Hu, Z. Feng, C. Hou, P. Rong, and W. Pang, ``SS-TBN: A Semi-Supervised Tri-Branch Network for
COVID-19 Screening and Lesion Segmentation,'' \emph{IEEE Trans. Pattern Anal. Mach. Intell.}, vol. 45, no. 8,  pp. 10427-10442, Aug. 2023.

\bibitem{b16} N. R. Blumenfeld \emph{et al}, ``Multiplexed reverse-transcriptase quantitative polymerase chain reaction using plasmonic nanoparticles for point-of-care COVID-19 diagnosis,'' \emph{Nat. Nanotechnol.}, vol. 17, no. 9,  pp. 984-992, Jul. 2022.

\bibitem{r9} R. W. Peeling \emph{et al}, ``Diagnostics for COVID-19: moving from pandemic response to control,'' \emph{The Lancet}, vol. 399, no. 10326, pp. 757-768, Feb. 2022.

\bibitem{b17} C. H. Woo, S. Jang, G. Shin, G. Y. Jung, and J. W. Lee, ``Sensitive fluorescence detection of SARS-CoV-2  RNA in clinical samples via one-pot isothermal  ligation and transcription,'' \emph{Nat. Biomed. Eng.}, vol. 4, no. 12,  pp. 1168-1179, Sep. 2020.

\bibitem{b18} L. C. Rosella \emph{et al}, ``Large-scale implementation of rapid antigen testing 
system for COVID-19 in workplaces,'' \emph{Sci. Adv.}, vol. 8, no. 8,  pp. 1-8, Feb. 2022.

\bibitem{b19} M. García-Fiñana  and I. E. Buchan, ``Rapid antigen testing in COVID-19 responses,'' \emph{Science}, vol. 372, no. 6542,  pp. 571-572, May 2021.

\bibitem{b20} P. Singh and Y. -P. Huang,  ``An Ambiguous Edge Detection Method for Computed Tomography Scans of Coronavirus Disease 2019 Cases,'' \emph{IEEE Trans. Syst., Man Cybern., Syst.}, vol. 54, no. 1,  pp. 352-364, Jan. 2024.


\bibitem{b22} S. Dong, Q. Yang, Y. Fu, M. Tian, and C. Zhuo, ``RCoNet: deformable mutual information maximization and high-order uncertainty-aware learning for robust COVID-19 detection,'' \emph{IEEE Trans. Neural Netw. Learn. Syst.}, vol. 32, no. 8,  pp. 3401-3411, Jun. 2021.

\bibitem{b23} R. Majumder, S. Jana, D. Ghose, ``Game theory-based decision making for the
allocation of scarce medical resources in the COVID-19 situation,'' \emph{IEEE Trans. Syst., Man Cybern., Syst.}, vol. 54, no. 4,  pp. 2137-2148, Apr. 2024.

\bibitem{b24} H. O. Tabrizi \emph{et al}, ``A low-cost handheld reconfigurable impedimetric
readout system for diagnostics of viral infections,'' \emph{IEEE Trans. Instrum. Meas.}, vol. 72, pp. 1-8, Apr. 2023.

\bibitem{b25} H. Wang \emph{et al}, ``Back-gate fully depleted silicon-on-insulator
P-channel Schottky barrier MOSFET with ultrahigh voltage sensitivity for label-free virus RNA detection,'' \emph{IEEE Trans. Instrum. Meas.}, vol. 73, pp. 1-8, Apr. 2024.

\bibitem{b26} A. Q. Rubio \emph{et al}, ``De novo design of modular and tunable 
protein biosensors,'' \emph{Nature}, vol. 591 , no. 7850, pp. 482-487, Jan. 2021.

\bibitem{b27} S. A. Perdomo \emph{et al}, ``SenSARS: A Low-cost portable electrochemical system for ultra-sensitive, near real-time,
diagnostics of SARS-CoV-2 infections,'' \emph{IEEE Trans. Instrum. Meas.}, vol. 70, pp. 1-10, Aug. 2021.

\bibitem{b28} A. T. Purnomo \emph{et al}, ``Non-contact supervision of COVID-19 breathing behaviour with FMCW radar and stacked ensemble learning model in real-time,'' \emph{IEEE Trans. Biomed. circuits Syst.}, vol. 16, no. 4, pp. 664-678, Aug. 2022.

\bibitem{b29} T. K. Dash, C. Chakraborty, S. Mahapatra, G. Panda, ``Gradient boosting machine and efficient
combination of features for speech-Based detection of COVID-19,'' \emph{IEEE J. Biomed. Health Inform.}, vol. 26, no. 11, pp. 5364-5371, Nov. 2022.

\bibitem{b30} H. Li \emph{et al}, ``Wireless, battery-free, multifunctional integrated bioelectronics for respiratory
pathogens monitoring and severity evaluation,'' \emph{Nat. Commun.}, vol. 14, no. 1, pp. 1-13, Nov. 2023.

\bibitem{b31} Y. Zhu, A. Tiwari, J. Monteiro, S. Kshirsagar and T. H. Falk, ``COVID-19 detection via fusion of modulation
spectrum and linear prediction speech features,'' \emph{IEEE Trans. Audio, Speech, Lang. Process.}, vol. 31, pp. 1536-1549, Apr. 2023.

\bibitem{b32} I. Aytekin \emph{et al}, ``COVID-19 Detection From Respiratory Sounds With Hierarchical Spectrogram Transformers,'' \emph{IEEE J. Biomed. Health Inform.}, vol. 28, no. 3, pp. 1273-1284, Mar. 2024.


\bibitem{b33} T. Dou, Z. Li, J. Zhang, A. Evilevitch, D. Kurouski, ``Nanoscale structural characterization of individual viral particles using atomic force microscopy infrared spectroscopy (AFM-IR) and tip-enhanced Raman spectroscopy (TERS),'' \emph{Anal. Chem.}, vol. 92, no. 16, pp. 11297-11304, Jul. 2022.


\bibitem{b35} S. Agarwal, R. Srivastava,  S. Kumar, Y. K. Prajapati, ``COVID-19 Detection Using Contemporary Biosensors and Machine Learning Approach: A Review,'' \emph{IEEE Trans. Nanobiosci.}, vol. 23, no. 2, pp. 291-299, Dec. 2024.

\bibitem{b36} A. Banerjee \emph{et al}, ``Rapid classification of COVID-19 severity by ATR-FTIR spectroscopy of plasma samples,'' \emph{Anal. Chem.}, vol. 93, no. 30, pp. 10391-10396, Jul. 2021.

\bibitem{c1} V. G. Barauna \emph{et al}, ``Ultrarapid On-Site Detection of SARS-CoV‑2 Infection Using Simple ATR-FTIR Spectroscopy and an Analysis Algorithm: High Sensitivity and Specificity,'' \emph{Anal. Chem.}, vol. 93, no. 5, pp. 2950-2958, Jan. 2021.

\bibitem{b37} I. Amenabar \emph{et al}, ``Structural analysis and mapping of individual
protein complexes by infrared nanospectroscopy,'' \emph{Nat. Commun.}, vol. 4, no. 1, pp. 1-9, Dec. 2013.


\bibitem{b38} J. Ye \emph{et al}, ``Accurate virus identiﬁcation with interpretable
Raman signatures by machine learning,'' \emph{Proc. Nat. Acad. Sci.}, vol. 119, no. 23, pp. 1-12, Jun. 2022.


\bibitem{b39} S. X. Leong \emph{et al}, ``Noninvasive and point-of-care surface-enhanced Raman scattering (SERS)-based breathalyzer for mass screening of coronavirus disease 2019 (COVID-19) under 5 min,'' \emph{ACS nano}, vol. 16, no. 2, pp. 2629-2639, Jan. 2022.


\bibitem{b41} D. Zhang \emph{et al}, ``Integrated system for on-site rapid and safe screening of COVID-
19,'' \emph{Anal. Chem.}, vol. 94, no. 40, pp. 13810-13819, Oct. 2022.

\bibitem{b42} Z. Movasaghi , S. Rehman, and I. U. Rehman, ``Fourier transform infrared (FTIR) spectroscopy of
biological tissues,'' \emph{Appl. Spectrosc. Rev.}, vol. 43, no. 2, pp. 134-179, Feb. 2008.

\bibitem{b43} M. H. C. Nascimento \emph{et al}, ``Noninvasive diagnostic for COVID-19 from saliva biofluid via FTIR spectroscopy and multivariate analysis,'' \emph{Anal. Chem.}, vol. 94, no. 5, pp. 2425-2433, Jan. 2022.

\bibitem{b44} B. R. Wood \emph{et al}, ``Infrared based saliva screening test for COVID-19,'' \emph{Angew. Chem.-Int. Edit.}, vol. 133, no. 31, pp. 17239-17244, Jun. 2021.

\bibitem{b45} L. Zhang \emph{et al}, ``Fast Screening and Primary Diagnosis of COVID-19 by ATR–FT-IR,'' \emph{Anal. Chem.}, vol. 93, no. 4, pp. 2191–2199, Feb. 2021.

\bibitem{b46} S. T. Kazmer \emph{et al}, ``Pathophysiological Response to SARS-CoV-2 Infection Detected by Infrared Spectroscopy Enables Rapid and Robust Saliva Screening for COVID-19,'' \emph{Biomedicines}, vol. 10, no. 2, p. 351, Feb. 2022.

\bibitem{d1} T. Mehmood, K. H. Liland, L. Snipen, and S. Sæbø, ``A review of variable selection methods in Partial Least Squares Regression,'' \emph{Chemom. Intell. Lab. Syst.}, vol. 118, pp. 62–69, Aug. 2012.

\bibitem{d2} A. C. Walls, Y.-J. Park, M. A. Tortorici, A. Wall, A. T. McGuire, and D. Veesler, ``Structure, Function, and Antigenicity of the SARS-CoV-2 Spike Glycoprotein,'' \emph{Cell}, vol. 181, no. 2, pp. 281-292.e6, Apr. 2020.

\bibitem{b47} P. Lahiri \emph{et al}, ``Fast Viral Diagnostics: FTIR-Based Identification, Strain-Typing, and Structural Characterization of SARS-CoV-2,'' \emph{Anal. Chem.}, vol. 96, no. 37, pp. 14749–14758, Sep. 2024.

\bibitem{c2} B. F. O. Coelho \emph{et al}, ``On the feasibility of Vis--NIR spectroscopy and machine learning for real time SARS-CoV-2 detection,'' \emph{Spectroc. Acta Pt. A-Molec. Biomolec. Spectr.}, vol. 308,  pp. 123735, Mar. 2024.

\bibitem{c3} M. Kamruzzaman \emph{et al}, ``Effect of variable selection algorithms on model performance for predicting moisture content in biological materials using spectral data,'' \emph{Anal. Chim. Acta}, vol. 1202,  pp. 339390, Apr. 2022.

\bibitem{c4} L. Sun \emph{et al}, ``Feature selection with missing labels using multilabel fuzzy neighborhood rough sets and maximum relevance minimum redundancy,'' \emph{IEEE Trans. Fuzzy Syst.}, vol. 30, no. 5,  pp. 1197-1211, May 2022.

\bibitem{c5} B. S. Zhang \emph{et al}, ``A novel random multi-subspace based ReliefF for feature selection,'' \emph{Knowledge-Based Syst.}, vol. 252,  pp. 109400, Sep. 2022.

\bibitem{c6} S. Mohdiwale \emph{et al}, ``Statistical wavelets with harmony searchBased optimal feature selection
of EEG signals for motor imagery classification,'' \emph{IEEE Sens. J.}, vol. 21, no. 13,  pp. 14263-14271, Jul. 2021.

\bibitem{s1}World Health Organization, ``Target product profiles for priority diagnostics to support response to the COVID-19 pandemic v. 1.0,'' \emph{}, Sep. 2020. [online]. 
Available:  https://www.who.int/publications/m/item/covid-19-target-product-profiles-for-priority-diagnostics-to-support-response-to-the-covid-19-pandemic-v.0.1. 

\bibitem{s2} W. Xing \emph{et al}, ``A Highly Automated Mobile Laboratory for On-site Molecular Diagnostics in the COVID-19 Pandemic,'' \emph{Clin. Chem.}, vol. 67, no. 4,  pp. 672-683, Apr. 2021.

\bibitem{s3} B. Diao \emph{et al}, ``Accuracy of a nucleocapsid protein antigen rapid test in the diagnosis of SARS-CoV-2 infection,'' \emph{Clin. Microbiol. Infec.}, vol. 27, no. 2,  pp. 289.e1-289.e4, Feb. 2021.

\bibitem{s34} J. Huang \emph{et al}, ``Rapid Detection of SARS-CoV-2 in Clinical and Environmental Samples via a Resonant Cavity SERS Platform within 20 min,'' \emph{ACS Appl. Mater. Interfaces}, vol. 15, no. 44, pp. 50742–50754, Nov. 2023.

\end{thebibliography}
\end{document}